\documentclass[manuscript,screen]{acmart}
\pdfoutput=1 
\usepackage{xcolor}
\usepackage{colortbl,booktabs} 
\usepackage{longtable}
\usepackage{multirow}
\usepackage{longtable}
\usepackage{bbding}
\usepackage{array}
\usepackage{hhline}
\newcolumntype{P}[1]{>{\centering\arraybackslash}p{#1}}

\usepackage{makecell}
\AtBeginDocument{
  \providecommand\BibTeX{{
    \normalfont B\kern-0.5em{\scshape i\kern-0.25em b}\kern-0.8em\TeX}}}

\setcopyright{acmcopyright}
\copyrightyear{2023}
\acmYear{2023}
\setcopyright{none}

\begin{document}

\title[Literature review]{Security for Machine Learning-based Software Systems: a survey of threats, practices and challenges}

\author{Huaming Chen}
\affiliation{
\institution{The University of Sydney}
\city{Sydney}
\country{Australia}
}
\email{huaming.chen@sydney.edu.au}
\author{M. Ali Babar}
\affiliation{
\institution{CREST - The Centre for Research on Engineering Software Technologies, The University of Adelaide}
\city{Adelaide}
\country{Australia}
}
\affiliation{\institution{Cyber Security Cooperative Research Centre}
\country{Australia}}
\email{ali.babar@adelaide.edu.au}

\renewcommand{\shortauthors}{Huaming Chen and M. Ali Babar}
\begin{abstract}
The rapid development of Machine Learning (ML) has demonstrated superior performance in many areas, such as computer vision, video and speech recognition. It has now been increasingly leveraged in software systems to automate the core tasks. However, how to securely develop the machine learning-based modern software systems (MLBSS) remains a big challenge, for which the insufficient consideration will largely limit its application in safety-critical domains. One concern is that the present MLBSS development tends to be rush, and the latent vulnerabilities and privacy issues exposed to external users and attackers will be largely neglected and hard to be identified. Additionally, machine learning-based software systems exhibit different liabilities towards novel vulnerabilities at different development stages from requirement analysis to system maintenance, due to its inherent limitations from the model and data and the external adversary capabilities. The successful generation of such intelligent systems will thus solicit dedicated efforts jointly from different research areas, i.e., software engineering, system security and machine learning. Most of the recent works regarding the security issues for ML have a strong focus on the data and models, which has brought adversarial attacks into consideration. In this work, we consider that security for machine learning-based software systems may arise from inherent system defects or external adversarial attacks, and the secure development practices should be taken throughout the whole lifecycle. While machine learning has become a new threat domain for existing software engineering practices, there is no such review work covering the topic. Overall, we present a holistic review regarding the security for MLBSS, which covers a systematic understanding from a structure review of three distinct aspects in terms of security threats. Moreover, it provides a thorough state-of-the-practice for MLBSS secure development. Finally, we summarise the literature for system security assurance, and motivate the future research directions with open challenges. We anticipate this work provides sufficient discussion and novel insights to incorporate system security engineering for future exploration.
\end{abstract}


\maketitle
\section{Introduction}
\label{sec:intro}
\textbf{Background of this work:} Machine learning models have dominantly outperformed many traditional statistical algorithms and models in various application areas, such as tasks from computer vision, video, speech and natural language recognition~\cite{wiriyathammabhum2016computer,zhang2020adversarial}, with an enhanced prediction and inference capability. Considering machine learning models as the core, one key research area is to improve the model performance dedicated to the systems both practically and theoretically with advanced functionalities, such as enhancing the object detection capability~\cite{boukerche2021visionbased}, human activity recognition~\cite{chen2021deeplearning}, boosting the recommendation service and facilitating the decision making process~\cite{zhang2019deep}. Typically, the success of such a machine learning-based software system requires tremendous efforts and dedicated resources by the people jointly from different research areas, e.g. machine learning, software engineering and data science. Similar to the classic software systems, the secure development of MLBSS is founded by the collaborative work regarding requirement engineering, framework/system/operation/process design, model development, system implementation and maintenance. A substantial interest has been attracted to address the development challenges for machine learning-based software systems, which predominantly focus on the aspects of \textit{technical debts} and \textit{quality assurance}.\par
\indent \textbf{Motivation of this work:} Recent studies have revealed that security threats for MLBSS could result in catastrophic consequences, particularly when it is targeted on safety-critical software systems. In 2020, the Software Engineering Institute at the CERT Coordination Center (SEI/CERT) has released a first documented vulnerability note for one common vulnerability and exposures (CVE) in a machine learning-based commercial software system~\cite{cert2020}. Another security-related system example is the supervisory control and data acquisition systems (SCADA), in which the intrusion detection system plays a vital role to protect SCADA from intrusion~\cite{suaboot2020taxonomy}. Supervised machine learning model is one pivotal component for the detection of intrusion to ensure the SCADA security. While machine learning models exhibit uncertainty and more attacking vectors are available for adversarial attacks, the security of machine learning-based safety-critical software systems becomes a major concern. One trending sub-topic is to secure the ML/DL models with robustness~\cite{biggio2018wild}. While recent works have centered on the security and defense methods for deep learning models~\cite{papernot2018marauder,he2020towards}, there will be a cat-and-mouse game.\par
\indent Thus, how to systematically secure MLBSS in practice is important and now a hot topic for the academia and industry. On the one side, the \textit{state-of-the-practice} has demonstrated a lack of diligence to ensure the deployment of machine learning-based systems following standardized processes, even with software testing techniques~\cite{serban2020adoption}. Most of the research focus has been on addressing the challenges in deploying the machine learning models in the practical applications from the software engineering point of view~\cite{liu2019secure,ashmore2021assuring}. For example, over 3,000 questions about DL software deployment to server/cloud/mobile/browser platforms from Stack Overflow were evaluated and a taxonomy consisting of 72 categories linked to challenges in the deployment was recently discussed~\cite{chen2020comprehensive}. The issues of efficiency and effectiveness in adopting methods and metrics from tradition software are reported to understand the challenges while lacking the practitioners face~\cite{amershi2019software,zhang2019empirical,zhang2019software,zhang2020machine}. The ethics, the end users' trust and the security issues as the cross-cutting aspects have been raised as general challenges hindering the scalability~\cite{paleyes2020challenges}. To alleviate the latent aspects, recently a framework named Machine Learning Technology Readiness Levels (MLTRL) framework is proposed to ensure the system operating in a robust, reliable and responsible manner~\cite{lavin2101technology,lavin2020technology}.\par
\indent On the other side, enabling the advanced MLBSS with current available practices will expose a even larger vulnerability surface while the consideration of security is not fully blended. Thus, more joint efforts from academia and industrial partners are desired to provide an overall view encompassing the identification of latent vulnerability root causes and the attack consequences understanding, to consolidate the system defence capabilities in MLBSS. One recent example is that, over 87\% Android apps based on TensorFlow Lite from Google Play are structurally similar to each other, which has resulted a successful adversarial attack on all 10 representative Android apps~\cite{huang2021robustness}. For edge devices, it requires more insights while deploying machine learning models in terms of the transferability of customised attack modes. Another example is from the system design level, for which the attacks will be easy to be propagated and transferred between systems with the convenience of open source ML/DL libraries and pre-trained models from distributed resources.\par
\indent \textbf{Our aim:} In this work, we aim to investigate the security of MLBSS with current secure development practices. System security engineering has been a long-term and critical research topic, whose primary objective is to `minimize or contain defence system vulnerabilities to known or postulated security threats and to ensure that developed systems protect against these threats.'~\cite{sebokwiki2021systemsecurity}. To achieve the goals of system security engineering, it is demanded to obtain a comprehensive understanding of security threats for MLBSS covering the novel attack modes and affecting development stages~\cite{nist2018systems}. Since the exploitation of vulnerability in MLBSS may occur at any stages, such as the requirement engineering and model development, how to distill the secure development practices is utterly important. For example, the internal defective design and incomplete secure practices for MLBSS could lead to the system failure by exposing the latent vulnerability to external dedicated attackers~\cite{serban2020adoption}. This happens when the practitioners could not transform the problem into requirement in an accurate and secure manner, which may bring undesirable implementation framework and process. Moreover, the vulnerability underlying the machine learning algorithms are considered as inherent limitations that could not be totally fixed yet. Thus, the external adversary with a dedicated level of machine learning knowledge will have the chance to attack the system either by poisoning malicious data or by the existing vulnerability transmitted from other ML models.\par
\indent To address this question, utilising the system security tools, including the threat model and CIA (confidentiality, integrity and availability) triad, for the integrated analysis of such system, will be firstly explored. In this work, one consolidated contribution is that we obtain the insights towards leveraging secure development practices for MLBSS at different software development stages for the emerging various attack modes. In details, we focus on the following research questions:\par
\begin{itemize}
    \item RQ1: What are the security threats in machine learning-based software systems? (Section~\ref{sec:taxonomy})
    \item RQ2: What are the state-of-the-art secure development practices in machine learning-based software systems? (Section~\ref{sec:security_practice})
    \item RQ3: What are the challenges and future directions for current practices in machine learning-based software systems? (Section~\ref{sec:challenges_futuredirections})
\end{itemize}

\section{Existing works and the survey structure}
\label{sec:structure}
\subsection{Existing works}
\begin{figure}[hb]
  \centering
  \includegraphics[width=0.7\linewidth,keepaspectratio]{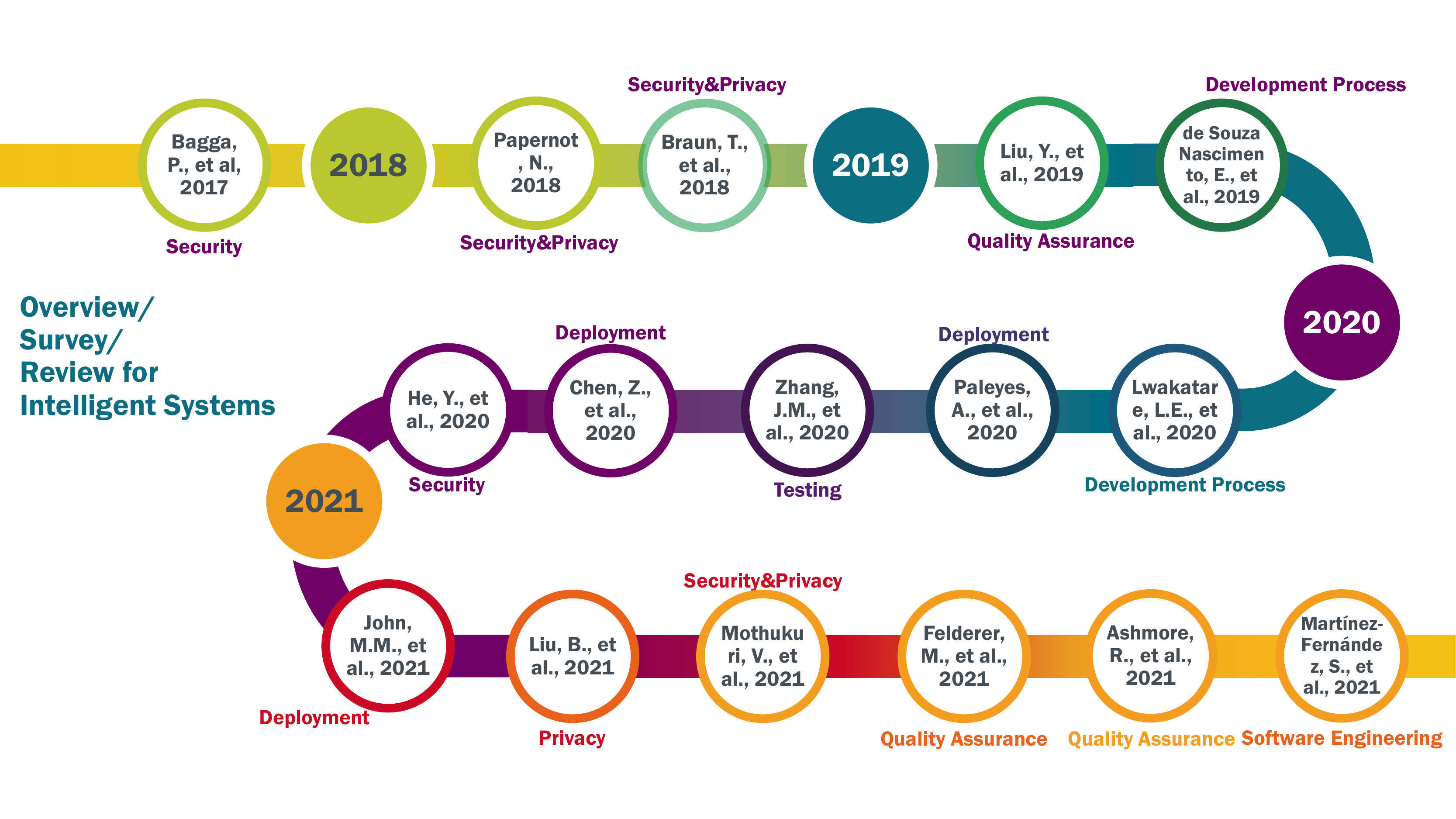}
  \caption{Existing work of overview/survey/review for intelligent systems.}
  \label{fig:existing_work}
\end{figure}

It is inspiring to witness the prominent research trend to utilising artificial intelligence techniques, particularly machine learning, in modern software systems. This brings an increasing number of surveys that summarise the works from different perspectives with the interest of a specific application domain, such as addressing the cloud computing security threats with different types of machine learning methods~\cite{butt2020review,nassif2021machine,gill2022ai}. A vast majority of these surveys has limited the scope either on a particular topic of the software engineering or a typical type of machine learning, which we have collectively depicted in Fig.~\ref{fig:existing_work}. Overall, there are three general categories of these works, which are \textit{`assurance'}, \textit{`software engineering practice'} and \textit{`security\&privacy'}. Table.~\ref{table:comparison} lists the comparison between our work and the existing work.\par
\indent The first group of existing review is related to the `assurance' of software systems. An exploring study for the quality assurance of deep learning based system was conducted by Liu et al.~\cite{liu2019secure}, initialising a discussion limited to deep learning based system engineering. While secure deep learning engineering is presented as a future approach for quality assurance, actionable items are not identified for the relevant process. An instant activity is referred to testing, one dominant methodology to capture the model behavior from a small group of studies. Zhang et al.~\cite{zhang2020machine} has thus surveyed the topic of machine learning testing to understand the different scenario for machine learning-based software systems. Recently, a comprehensive survey for the assurance of machine learning, i.e., its lifecycle and the generation of assurance evidence, is presented by Ashmore et al.~\cite{ashmore2021assuring}. It has generally discussed the machine learning errors. Meanwhile, the methods which provide evidence to support ML components fitting for the purpose and could be adequately integrated into the systems are also included. The key point is raised that the assurance should cover all stages of ML lifecycle. Another recent work by Felderer et al.~\cite{felderer2021quality} has highlighted the challenges on quality assurance encountered in the intersection between software engineering and artificial intelligence for artificial intelligence-based systems, which has exhibited a control transfer from typical source code to data.\par
\indent Another group of existing survey has a focus on the `software engineering practices' (SE practices) for machine learning-based software systems. The work by Mart$\acute{i}$nez-Fern$\acute{a}$ndez et al.~\cite{martinez2021software} has explicitly discussed the definition of artificial intelligence-based systems, which core functionalities are enabled by at least one AI component (e.g., for image- and speech-recognition, and autonomous driving). It targets the topic of `software engineering for artificial intelligence' (SE4AI) by reviewing the works that explore software engineering practices to develop, maintain and evolve AI-based systems. Another work by Lwakatare et al.~\cite{lwakatare2020large} is a review work that investigated the well-established development guidelines for traditional software systems in the scenario of industrial settings to develop and maintain the large-scale machine learning-based software systems. Specifically, there is a plethora of works investigating the deployment process of machine learning-based software systems. Chen et al.~\cite{chen2020comprehensive} has conducted a survey that mined and analysed 3,023 posts from public developer forums to identify the encountered specific challenges in the process of deep learning-based software systems deployment. Other works have either utilised systematic literature review method~\cite{john2020architecting,paleyes2020challenges,giray2021software} or conducted interviews with developers~\cite{de2019understanding} to understand the challenges and solutions in deploying machine learning-based software systems.\par
\indent The last group targets at `security\&privacy'. Begga et al.~\cite{bagga2017mobile} and Braun et al.~\cite{braun2018security} are the earliest works that anticipate to understand the security and privacy threats in intelligent systems, in which they focus on mobile agent-based applications and smart cities, respectively. While it is a general review, Papernot et al.~\cite{papernot2018marauder} initialised the discussion of security and privacy issues for machine learning-based software systems by exploring the threat model spaces for the learned models. He et al.~\cite{he2020towards} is so far the first work that review the potential attack modes towards deep learning-based systems, in which four types of attacks were summarised. Privacy preservation problem for machine learning is surveyed by Liu et al.~\cite{liu2021machine}. Mothukuri et al.~\cite{mothukuri2021survey} presented that the security and privacy issues persist and will demand further investigation and collaboration for federated machine learning. However, these works have limited the discussion to survey security threats from the algorithmic perspective.\par
\subsection{Comparison with other works}
\indent While it is similar to the development of traditional Internet infrastructure and system, the machine learning-based software systems are more complicated by the fact that it currently lacks in-depth understanding of the decision-making process. In Table.~\ref{table:comparison}, a comparison with existing works is presented. It is noted that, with the definition of SE practices, some works have focused on the discussion of software quality assurance~\cite{felderer2021quality,liu2019secure}. A particular interest of providing evidence as assurance to support MLBSS is largely surveyed in~\cite{ashmore2021assuring}. While these works are dedicated to assuring the MLBSS behaviour, the fundamental security development practice is missing. In the work of~\cite{zhang2020machine}, machine learning model testing, as a key method to mitigate the security threats, is broadly discussed from the perspective of testing process. Other works incorporating the SE practice, such as~\cite{martinez2021software,john2020architecting,lwakatare2020large}, have missed the discussion of security threats in MLBSS. For the rest works, they have demonstrated a similar research topic related to the security. Some have identified the distinct attack types and privacy issues~\cite{he2020towards,braun2018security,bagga2017mobile,liu2021machine}, such as adversarial attack~\cite{sun2018adversarial,chakraborty2018adversarial}, poisoning attack~\cite{goldblum2020dataset}, and membership inference attack~\cite{hu2022membership} and so on. In summary, a holistic understanding of security threats for the emerging various attacks is missing. \par
\indent A most related work by~\cite{papernot2018sok} has been published in 2018 to discuss the attacks and defences of ML systems. It is motivated to see the highlights of CIA model for attack surfaces characteristics and the consideration of machine learning pipeline for adversarial goal identification. While it has aimed at a theoretical understanding of the sensitivity of machine learning models, it has successfully attracted a huge amount of research to the machine learning attacks and defences. However, the scope is limited to the structural elements of \textit{machine learning algorithms} and the \textit{data used to train them}, which has now been outdated. {Overall, there is no literature review that is aimed at a comprehensive investigation of machine learning-based software systems from the secure development aspect, particularly with the understanding of systematic analysis of security threats and security engineering practices.\par
\indent Since the interference of such intelligent systems could occur at many stages in the virtual and physical world~\cite{herpig2019securing}, there is a need to thoroughly investigate the secure development practices for the machine learning-based software systems~\cite{mcgraw2019security,schneier2020attacking} with the lens of system security. Thus, this work will \textbf{firstly} contribute an exclusive taxonomy of attack, which also includes the discussion of details for the identified attack vectors, the impacts and potential defence strategies as the \textbf{second contribution}. Most importantly, it advocates this work to capture the state-of-the-practices for MLBSS system security as the \textbf{third contribution}. \textbf{Lastly}, it provides a conjunction discussion of security practices, which will help us to identify the limitations and future research directions.\par
\begin{table}
\begin{tabular}{|p{3cm}|p{2.5cm}|p{2.5cm}|P{1.3cm}|P{2cm}|P{2cm}|}\hline
\textbf{Study} & \textbf{Subtopics} & \textbf{Domain} & \textbf{SE practice} & \textbf{System security} & \textbf{Secure development practice}\\\hline
Mart$\acute{i}$nez-Fern$\acute{a}$ndez et al. 2021~\cite{martinez2021software} & SE practices & AI-based systems & \Checkmark & \XSolid & \XSolid \\\hline
Asmore et al. 2021~\cite{ashmore2021assuring} & Quality Assurance & ML lifecycle & \XSolid & \Checkmark & \XSolid \\\hline
Felderer et al. 2021~\cite{felderer2021quality} & Quality assurance & AI-based systems & \Checkmark & \XSolid & \XSolid \\\hline
Liu et al. 2021~\cite{liu2021machine} & Security\&Privacy & Machine learning & \XSolid & \XSolid & \XSolid \\\hline
He et al. 2020~\cite{he2020towards} & Security\&Privacy & Deep learning systems & \XSolid & \Checkmark & \XSolid \\\hline
John et al. 2020~\cite{john2020architecting} & SE practices & Edge/cloud/hybrid architectures & \Checkmark & \XSolid & \XSolid \\\hline
Zhang et al. 2020~\cite{zhang2020machine} & SE practices & ML system & \XSolid & \Checkmark & \XSolid \\\hline
Lwakatare et al. 2020~\cite{lwakatare2020large} & SE practices & Industrial software systems& \Checkmark & \XSolid & \XSolid \\\hline
Liu et al. 2019~\cite{liu2019secure} & Quality assurance & Deep learning systems & \XSolid & \Checkmark & \XSolid \\\hline
Papernot et al. 2018~\cite{papernot2018sok} & Security\&Privacy& ML-based systems & \Checkmark & \Checkmark & \XSolid \\\hline
Braun et al. 2018~\cite{braun2018security} & Security\&Privacy& Smart cities & \XSolid & \Checkmark & \XSolid \\\hline
Bagga et al. 2017~\cite{bagga2017mobile} & Security\&Privacy & Mobile agent-based applications & \XSolid & \Checkmark & \XSolid \\\hline
Our work & Security and secure development practices & ML-based software systems & \Checkmark  & \Checkmark & \Checkmark\\\hline
\end{tabular}
\caption{Comparison between our work and the existing work related to intelligent systems}
\label{table:comparison}
\end{table}
\subsection{Methodology}
We have implemented this work as a structured literature review, with an inspiration of using broadly defined keyword search string for relevant literature identification. We firstly designed the search strings: `` (`vulnerability' OR `privacy' OR `attack') AND (`machine' OR `deep' OR `reinforcement' OR `federated') AND `learning based system' '', `` (`vulnerability' OR `privacy' OR `attack') AND (`artificial intelligence system')'', `` `security' AND (`machine' OR `deep' OR `reinforcement' OR `federated') AND `learning based system' ''. We aims to cover as many impactful paper as possible by soliciting the results from the sources from DBLP, Google Scholar, web in general formats such as papers, patents, courses, slides, tutorials, reports, articles and so on. After reading the titles and the abstracts, together with the backward and forward snowballing, we have included 141 papers for our survey. Particularly, we have identified 52 paper related to the security threats for MLBSS covering data- (15 papers), model- (30 papers), and system-oriented (7 papers) attacks. We wish not to claim that all the papers in this are collected, especially when this area is emerging in such an exceptional speed. However, we believe that our data collection process has covered most of the key studies to provide the security issues and practice for MLBSS. \par
\subsection{Overall structure of the work}
\indent Following, with the collected state-of-the-art literature works, we present the taxonomy of the latest security threats and the corresponding secure development practices in relation to different stages. The sections are organised as follows: section ~\ref{sec:taxonomy} presents a taxonomy of security threats in MLBSS as well as the attack types. Section ~\ref{sec:security_practice} highlights the state-of-the-practice for MLBSS and the limitations of adopting current practices to secure MLBSS. The open challenges and future directions are presented in section ~\ref{sec:challenges_futuredirections} while section ~\ref{sec:conclusion} concludes the paper.\par
\section{Attack taxonomy for system security in MLBSS}
\label{sec:taxonomy}
The recent immense success of machine learning-based software systems is not possible without the accumulating big data and cutting edge machine learning techniques~\cite{marx2013big}. For MLBSS, the advancement of machine learning models, particularly for the deep learning models with over hundred thousands parameters, is the critical contributor to learn the latent deep relationships from the big data to generate the outstanding prediction, inference as well as the reasoning performance in comparison to human intelligence. Moreover, it will not be possible without the development of computational infrastructure. Including the high performance computing cluster and the graphics processing unit (GPU), the supporting infrastructure for MLBSS has now become more complicated by including more advanced computational resources to support the model training and system deployment. These contributing factors are briefly discussed in~\cite{mothukuri2021survey}, and the recent representative applications include the AlphaFold 2 for science from DeepMind~\cite{jumper2021highly} and various mobile applications with on-device model for Android applications~\cite{huang2021robustness}. These factors have posed novel security threats, and new attack surfaces can be leveraged by attackers in either a proactive or reactive manner.\par
\indent The security threats in empowered software systems have recently aroused attention from different areas, including the general AI and cybersecurity communities. Nonetheless, a dominant focus has been limited to the aspects of algorithms and data. Via the systematic analysis of machine learning-based software systems, we first discuss the system security threats with confidentiality, availability and integrity (CIA) triad from the perspective of information security practitioners and researchers~\cite{samonas2014cia,beckers2015pattern,papernot2016towards}.\par
\indent 1. Confidentiality: it indicates the sensitive information and private data are not available or disclosed to unauthorised access, including individuals, entities and processes.\par
\indent 2. Integrity: it means the completeness and trustworthiness of data over the entire data lifecycle in system.\par
\indent 3. Availability: it guarantees that the information service are always online and available as intended at all times.\par
\indent While CIA triad could serve as good metrics for analyzing the security issues for MLBSS, it is originally designed for a general course of security analysis. For example, the confidentiality for traditional software systems could be digital data compromised, leakage and theft such as hardware theft, password theft and phishing emails, while the counter examples for availability includes the period of power outages, hardware failures, network interruptions and system upgrades. In following sections, we discuss the security threats of MLBSS from the system aspects of confidentiality, integrity and availability.\par
\subsection{Threat Model for MLBSS}
To design a secure software system, a list of system security questions can be utilised, in which one prominent structured approaches is `threat modelling'. Threat modelling demands a thorough consideration of `who', `what' and `how', which could structurally represent the available information for security analysis. Considering the MLBSS as a specific class of software systems, the questions will focus on who might want to abuse the system, what capabilities they will have, how they might be able to do it and what the overall risk of the exploitation is. System designers and engineers will be acknowledged with the output about where defence efforts should be focused and what compromises can be anticipated.\par
\indent Fig.~\ref{fig:mlbss_process} illustrates the five stages of the MLBSS lifecycle, which is extended from the machine learning workflow~\cite{amershi2019software}, covering the stages of problem definition, data management, model construction, system deployment and system maintenance. \par
\indent While traditional cyber attacks may still be effective targeting on MLBSS, we aim to cover the machine learning-based components and processes as the novel attack surfaces, in which attackers can utilise the model and supporting infrastructure defects affecting both software and hardware components at different development stages. The ultimate goal is to manipulate data, model and system. In comparison to the existing literature reviews, this work considers the security threats from a holistic view to analyse the overall system. It thus covers the security vulnerabilities from data attack, model attack and system attack. Moreover, the correlation among different types of attacks is investigated by providing the engineering practices discussion shown in Fig.~\ref{fig:mlbss_process}. It presents a broader and deeper discussion from the the system security point of view, which can provide a clear taxonomy to better serve the system design, testing and auditing goals. Fig.~\ref{fig:mlbss_taxonomy} illustrates the current taxonomy for attack types of MLBSS.\par
\begin{figure}[htbp]
  \centering
  \includegraphics[width=0.8\linewidth,keepaspectratio]{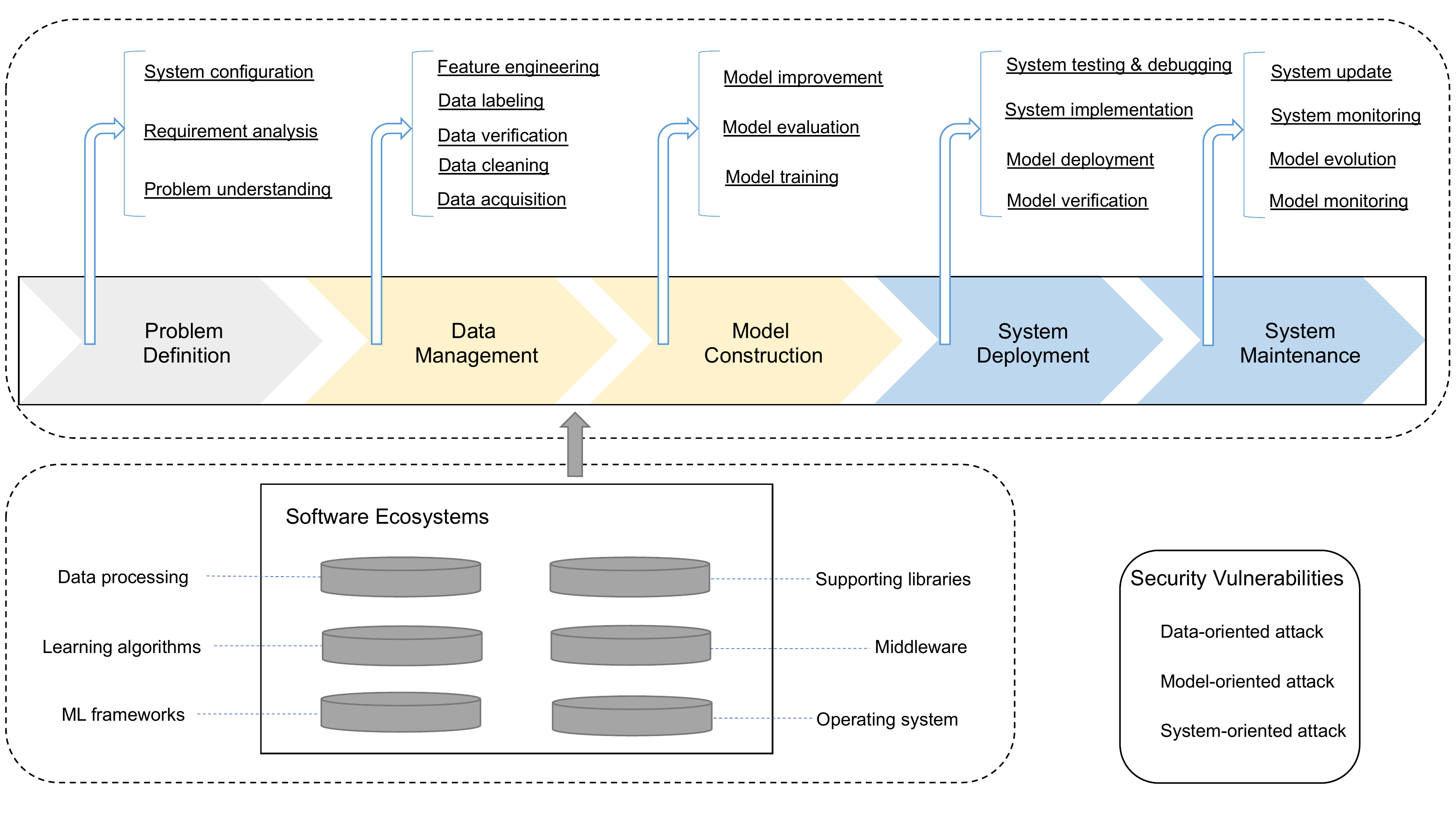}
  \caption{The five stages of the machine learning-based software system lifecycle}
  \label{fig:mlbss_process}
\end{figure}
\begin{figure}[htbp]
  \centering
  \includegraphics[width=0.8\linewidth,keepaspectratio]{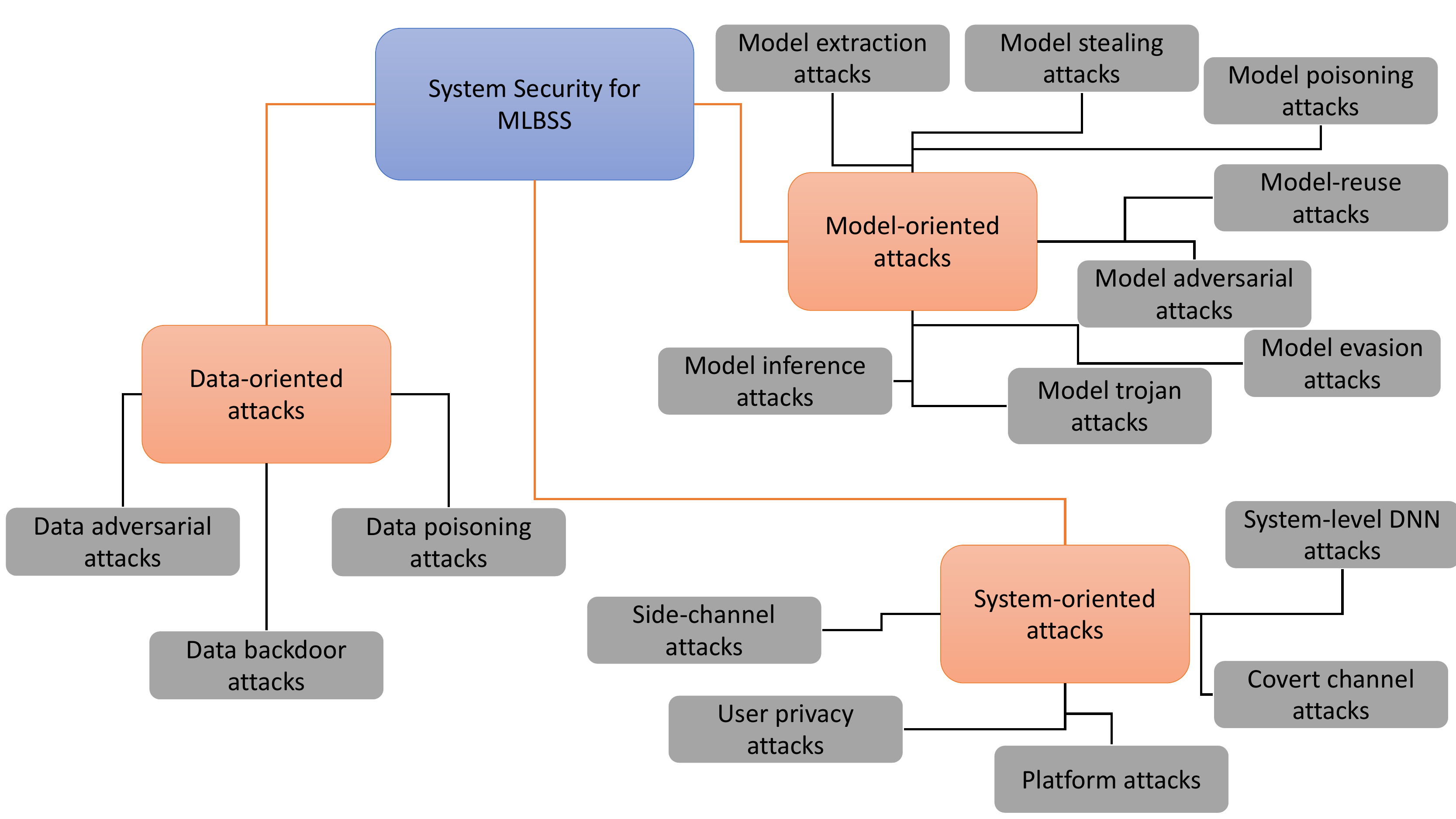}
  \caption{The taxonomy for system security of machine learning-based software systems}
  \label{fig:mlbss_taxonomy}
\end{figure}
\subsection{Data-oriented attacks}\label{subsec:data_attacks}
As per the phases in Fig.~\ref{fig:mlbss_process}, data is one of the pivotal assets in MLBSS, as well as the most concerned asset to security threats. Different from traditional software systems, data in MLBSS mostly refer to the training and serving data, whose quality is significant to the system performance. This is also a recently hot topic in modern data-driven systems, such as in ~\cite{fariha2021conformance} the successful deployment and accurate operation of system depend on the continuing conformance of the data to its initial design. Learning and extracting the latent relationships among the large amount of data is the general process of machine learning models. As an accessible vulnerable asset in system, \textbf{mislabelled and inferior data} is considered as one passive factor that impact the machine learning model security~\cite{he2020towards}. On the other hand, attackers can leverage different mechanisms to \textbf{access} the training data, \textbf{poison} the data, even \textbf{inject} the Trojan data.\par
\indent We measure the data-oriented attacks with respect to the adversary knowledge levels and adversarial goals that may directly impact the training and serving data. In summary, the main training processes include the offline settings (traditional learning on the collected data) and the online settings (such as active learning), which can be achieved by either a centralised way or distributed way. Active learning is designed for achieving higher performance with fewer data samples by querying the labels from an trusted oracle, e.g., an expert annotator~\cite{settles2009active}. For the latter distributed way, a notion is formalised as federated machine learning, which trains the model with multiple decentralised computational holding local data samples~\cite{yang2019federated}. The settings have a larger liability of exposing the attack surface to the adversary, in which the adversary can manipulate the training data in different stages. In general, the success of machine learning algorithms assumes the integrity of the training data and serving data. Availability of the training data is one main concern for security. The goals of data-oriented attacks can be summarised as to: 1. degrade the machine learning model performance; 2. manipulate the prediction outcome at test time.\par
\noindent \textbf{Review \& discussion:}\par
Overall, the data-oriented attacks could be categorised into three major types of attacks, which are poisoning attacks, backdoor attacks and adversarial attacks. While poisoning attacks have been a general term summarising most popular attacks, in this work poisoning attacks represents the triggerless poisoning attacks while the backdoor attacks is the backdoor poisoning attacks. Triggerless poisoning attacks refer to successfully mislead the model and system behaviour without manipulating of the data during model inference stage. For backdoor attacks, the attack will be activated when the input data contains crafted trigger, which could be pixel-level features in an image or designed characters in a sequence. In Table.~\ref{tab:data_attack}, we have collectively summarised 15 most recently exemplar studies to cover the modes targeting on different development stages. \par
\indent For both triggerless poisoning attacks and backdoor poisoning attacks, the attacks mostly happen during the data management stage when the attackers could easily obtain the access to the dataset. This assumption is made against the centralised machine learning models~\cite{suciu2018does,jagielski2018manipulating,wang2021robust,schwarzschild2021just,barni2019new,xiang2020detection,li2020rethinking}, whilst one specific scenario is identified for the distributed machine learning models. In ~\cite{tolpegin2020data,nuding2020poisoning}, the federated learning-based systems are studied that the global model is poisoned by aggregating the edge node model updates learned from malicious participants. The last attacking stage for backdoor poisoning attacks is discovered in ~\cite{li2021backdoor}, in which the model is retrained periodically at system maintenance stage. Generally, attacks have impacts on the data integrity and availability.\par
\indent Although several earlier works~\cite{perdisci2006misleading,nelson2008exploiting,rubinstein2009antidote,mozaffari2014systematic} have identified the poisoning attacks in different applications, one earliest work presenting a more realistic assumption by limiting the adversary's capability and knowledge of the system was led by Suciu et al.~\cite{suciu2018does}. Neither the feature nor the algorithm knowledge was taken for granted as the adversary is limited with four constraints to generate the poison samples in a more inconspicuous way. StingRay attack was proposed as a \textit{model agnostic black-box poisoning attack} against four different machine learning-based software systems for three classification models. In another work from Jagielski et al.~\cite{jagielski2018manipulating}, linear regression learning models, one of the efficient algorithms in machine learning, were firstly studied against \textit{poisoning availability attacks}. Two attack methodologies, \textit{optimization-based poisoning attacks (OptP)} and \textit{statistical-based poisoning attacks (StatP)}, were developed, while one novel defence algorithm named \textit{TRIM} was evaluated on three public regression datasets. Federated machine learning is one emerging paradigm which benefits from its design of distributed learning of large scale deep learning model and large amount of data. However, federated machine learning also introduces new vulnerabilities while providing user privacy. The AI chat bot \textit{Tay} by Microsoft was reported to be manipulated by malicious users from its online supervised learning model to learn offensive and racist language~\cite{schlesinger2018let}. In~\cite{tolpegin2020data}, the role of malicious participant contributing to federated machine learning model construction was discussed in the specific label flipping attack, in which the global model was poisoned by the malicious model updates.\par
\indent In order to understand the threat of poisoning attacks, Schwarzschild et. al~\cite{schwarzschild2021just} has recently developed a benchmarking framework to evaluate a wide range of poison attacks on machine learning models for images. Correspondingly, another work by Wang et al.~\cite{wang2021robust} studied the robustness of stochastic gradient descent to data poisoning attacks in a two-layer neural network. Both works have identified that the machine learning models optimised with stochastic gradient descent (SGD) could be significantly harder to poison.\par
\indent Backdoor poisoning attack, another type of \textit{poisoning integrity attack} aiming to manipulate the training data to influence the prediction outcome at test time, has emerged as a new form of data poisoning attacks. Sometimes, it is identified as `Trojan attacks'. Similar to other data poisoning attacks, attackers will have access to the training data. Adding the stealthy trigger is the main approach, i.e., local patch in the image classification case. One example was studied in Barni et al.~\cite{barni2019new}. Specifically, a more stealthier way by target data corruption without label poisoning was proposed and evaluated against convolution neural network (CNN) model. In the work by Xiang et al.~\cite{xiang2020detection}, defence strategy was evaluated given a trained DNN which may use a possibly backdoor-poisoned training data for learning. While data could be distributed in different computing nodes for federated machine learning either vertically or horizontally, Nuding et al.~\cite{nuding2020poisoning} studied the backdoor attacks from two different strategies of \textit{returning model learned from malicious node} and \textit{model replacement from backdoor node} to influence the final model.\par
\indent Another major type of data-oriented attacks is adversarial attack. The primary goal is to degrade run-time performance when systems are deployed. Thus, how to generate the malicious serving data to yield misleading results while remaining imperceptible is the key. For adversarial attacks, limiting the capability of the adversary to access the information of systems in a black-box setting is a more practical way when attacking system. One early work by Papernot et al.~\cite{papernot2017practical} studied to craft adversarial examples by a surrogate training model using synthetic data generation. Attacks were deployed against Amazon Machine Learning and Google's Cloud Prediction API platforms. Ilyas et al.~\cite{ilyas2018black} characterised three more restrictive scenarios. The threat models were defined with query-limited setting, partial-information setting and label-only setting. The novel adversarial attacks were successfully evaluated against a commercial system, which is the Google Cloud Vision (GCV) classifier. While many works focus on image-oriented systems, Boucher et al.~\cite{boucher2021bad} recently investigate vulnerability in text-based natural language processing (NLP) systems. For the deployed commercial systems from Microsoft, Google, Facebook and IBM, they have identified that a thorough input sanitization against adversarial attacks is demanded. A most recent adversarial attack example is the physical-world attack investigated by Sato et al.~\cite{sato2021dirty}. A \textit{Dirty Road Patch} attack was devised on the real-world automated lane centering (ALC) system, which was found to be robust to various real-world factors and general to multiple model design. Meanwhile, it is also stealthy from the driver's view.\par
\begin{table}
\centering
\fontsize{8}{9}\selectfont
\caption{Summary of the surveyed works for the \textit{Data-oriented attacks} for MLBSS.}
\label{tab:data_attack}
\begin{tabular}{|p{1.5cm}|p{2.5cm}|p{2.5cm}|p{2.5cm}|p{1.8cm}|p{2cm}|}
  \hline 
  \textbf{Study} & \textbf{Attack mode} & \textbf{Attack stage} & \textbf{Model} & \textbf{Affecting CIA} & \textbf{Defence}\\\hline
  Suciu et al. 2018~\cite{suciu2018does} & \multirowcell{6}[0ex][l]{\textit{Poisoning attack}} & Data management & CNN, linear SVM and random forest&  Data integrity & None\\\hhline{-~*{4}{-}}
  Jagielski et al. 2018~\cite{jagielski2018manipulating}& & Data management & OLS, Ridge regression, LASSO and Elastic-net regression& Data availability&TRIM algorithm\\\hhline{-~*{4}{-}}
  Tolpegin et al. 2020~\cite{tolpegin2020data}& & Model construction & CNN & Data \& Model integrity & Identify malicious model updates\\\hhline{-~*{4}{-}}
  Wang et al. 2021~\cite{wang2021robust}& & Data management & Neural Network & Data integrity & SGD \\\hhline{-~*{4}{-}}
  Schwarzschild et al. 2021~\cite{schwarzschild2021just}& & Data management & CNN & Data integrity & None\\\hline

  Barni et al. 2019~\cite{barni2019new}& \multirowcell{5}[0ex][l]{\textit{Backdoor attack}} & Data management & CNN & Data integrity & None\\\hhline{-~*{4}{-}}
  Xiang et al.~\cite{xiang2020detection}& & Data management & DNN & Data integrity & Unsupervised anomaly detection defence\\\hhline{-~*{4}{-}}
  Nuding et al. 2020~\cite{nuding2020poisoning}& & Model construction & CNN & Model integrity & None\\\hhline{-~*{4}{-}}
  Li et al. 2020~\cite{li2020rethinking} & & Data management & CNN & Data integrity & Transformation-based defence\\\hhline{-~*{4}{-}}
  Li et al. 2021~\cite{li2021backdoor} & & System maintenance & DNN & Data integrity & None \\\hline
  
  Papernot et al. 2017~\cite{papernot2017practical}& \multirowcell{3}[0ex][l]{\textit{Adversarial attacks}} & System deployment & DNN, logistic regression, SVM, decision tree and nearest neighbors & Data integrity & None \\\hhline{-~*{4}{-}}
  Ilyas et al. 2018~\cite{ilyas2018black}&  & System deployment & CNN & Data integrity & None\\\hhline{-~*{4}{-}}
  Boucher et al. 2021~\cite{boucher2021bad} &  & System deployment & Fairseq~\cite{ott2019fairseq}, Toxic Content Classifier\footnote{https://github.com/IBM/MAX-Toxic-Comment-Classifie}, Google's Perspective API\footnote{perspectiveapi.com}, RoBERTa~\cite{liu2019roberta} & Model integrity \& availability & None \\\hhline{-~*{4}{-}}
  Sato et al. 20201~\cite{sato2021dirty} & & System deployment & DNN & Data integrity & None\\\hline
\end{tabular}
\end{table}
\subsection{Model-oriented attacks}\label{subsec:model_attacks}
We consider the model-oriented attacks with respect to the adversary knowledge level and adversarial goals that can directly impact the learning models during the training and run-time without accessing the training data. In summary, the adversary will have access to: 1.) modify the training processes, 2.) manipulate the deployed model. To modify the training processes, the adversary has several options for different development stages. For model construction stage, the distribution of the pre-trained malicious models and the modification of the objective function are two main methods. An adversary can also inject security vulnerability during system deployment by adding model patches~\cite{kurita2020weight,costales2020live}. The other conventional attacks to manipulate the deployment model will be related to the privacy information leakage, in which the adversary targets on the extraction and inversion of the systems information. The adversary will target data memberships or properties information for model inversion attacks. In terms of model extraction attacks, the primary goal is to duplicate the deployed machine learning models. The inversion attacks can be launched in a white-box and black-box setting, while the extraction attacks mostly assume a black-box setting.\par
\textbf{Review \& discussion:}\par
In summary, the model-oriented attacks are categorised into seven types with different goals, which are neural Trojan attacks, model poisoning attacks, model inference attacks, model adversarial attacks, model reuse attacks, model stealing attacks and model extraction attacks (Fig.~\ref{fig:mlbss_taxonomy}). In table.~\ref{tab:model_attack}, we have collectively listed 30 most related studies to demonstrate the attack types for different development stages.\par
\indent For model poisoning attacks, the primary goal is to undermine the expected performance thus leading to the unusability of trained model. It normally happens in the model construction stage. Both model integrity and availability will be affected. One way to inject the vulnerabilities to the machine learning models is via the distributed pre-trained models. Kurita et al.~\cite{kurita2020weight} studied the situation that an attacker has no knowledge of the fine-tuning details but could initialise attacks by accessing the public datasets, which is realistic practice. The model can be exploited even after fine-tuning for the specified task. Another scenario is the distributed machine learning models. When the industry scale MLBSS has now become distributed, a robust distributed machine learning-based system will require Byzantine failure tolerance~\cite{blanchard2017machine,mahloujifar2019universal}. An attack will affect the model availability, as the model will end up with denial-of-service attacks, particularly for federated machine learning-based systems~\cite{fang2020local}. Also, the model poisoning attacks can threaten the learning process directly. In ~\cite{bhagoji2019analyzing}, the stealthy model poisoning attack was launched by optimising the malicious objective which is designed for targeted misclassification. While it showed that simply poisoning the data label may not be effective for federated machine learning, the model poisoning attacks appear to be more challenging since two different Byzantine-resilient aggregation mechanisms can not ensure that models will sustain a good performance under attack. Recently, Lorenz et al. studied two direct and indirect poisoning attacks towards deep neural network, which have either modified the training process or poisoned the data~\cite{lorenz2021backdoor}. The evaluation of the network certification robustness against training time attacks indicates that better defence methods are required.\par
\indent For neural Trojan attacks, it also refers to the model poisoning attack but it impacts the model behaviour at run-time when the trigger pattern specified by the attacker is presented. Its wide concerns have been raised due to the available pre-training process could not ensure its integrity. Either the deployed systems are supported by a ML model intellectual property vendor or the malicious patches on deep learning models, its performance on the specific trigger patterns will be largely degraded. While data-oriented attacks have an assumption that attackers have access to modify the training data directly, one type of neural Trojan attacks envisions the attackers benefit from the machine learning model outsourced training, such as the machine-learning-as-a-service provider~\cite{liu2017neural,liu2020survey}. Since current neural networks have become larger and bigger, the neural Trojan attacks have become pervasive. Liu et al.~\cite{liu2017neural} demonstrated the security threat of neural Trojan embedded in a neural network intellectual property (IP), which was trained by an IP vendor. Three techniques were proposed to mitigate the identified attacks. The security implications of using third-party primitive models were investigated in ~\cite{ji2018model}. A broad class of model-reuse attacks was demonstrated with four distinct machine learning-based systems from the areas of skin cancer screening, speech recognition, face verification and autonomous driving~\cite{ji2018model}. Feature extractors were leveraged to trigger the host systems to malfunction. Gu et al.~\cite{gu2019badnets} studied the backdoor attack by modelling two different attack scenarios, one was via fully outsourced training task and another is by maliciously pre-trained model from the Caffe Model Zoo and Keras Pre-trained Model Library. The increasingly general practices of outsourcing the training task due to the high demand computational resource and acquiring the pre-trained models from publicly available machine learning model zoo have brought new types of attack surface. It has now become a critical topic to guarantee the model integrity to adopt novel security protocols accordingly. Chen et al.~\cite{chen2019deepinspect} studied the problem of detecting whether a provided deep neural networks (DNN) model has been trojaned. One defence method name `DeepInspect' was developed to learn the probability distribution of the embedded triggers for the queried models. Another approach `TABOR' developed by Guo et al.~\cite{guo2019tabor} targeted different sizes, shapes and locations of triggers pertaining to the trojan and utilised the optimization regularisation technique to better detect trojans. One most recent study by Costales et al. has introduced another type of new live Trojan attack that can directly patch the model parameters at run-time~\cite{costales2020live}. By minimising the number of overwrites of model parameters in memory, the new live Trojan attack is considered as one of the stealthiest way to bypass the existing intrusion detection systems.\par
\indent Different from manipulating the MLBSS behaviour at run-time, other two types of model-oriented attacks are model inversion attacks and model extraction attacks. They share same characteristic to proactively extract the privacy information from systems. For model inversion attacks, ~\cite{ateniese2015hacking} is one of the earliest works that identified the training dataset leakage. The evaluation on an Internet traffic classifier based on SVM and a speech recognition software based on Hidden Markov models illustrate that, the release of trained machine learning model to the public is not safe. It may result in the information leakage of training dataset. Particularly, the defence method of differential privacy was proved to be ineffective in the scenario. In ~\cite{shokri2017membership}, a shadow training technique is devised to initialise the membership inference attacks against the black-box models training using Amazon ML and Google Prediction API. Shokri et al. have focused on determining whether one acquired data sample was used for the model training. As a mitigation strategy, the regularisation approach is evaluated to be necessary and useful. Salem et al.~\cite{salem2019ml} extended the work by relaxing the assumptions from ~\cite{shokri2017membership}, which are establishing shadow models with same structure as the target model and using a shadow training dataset with same distribution as target training dataset. The result shows that the membership inference attacks can be easily achieved with fewer limitations. These two studies have all applied to machine-learning-as-a-service (MLaaS) providers. One of the work for the membership inference attacks against federated machine learning-based system, or earlier termed as distributed deep learning-based system, was by Hitaj et al.~\cite{hitaj2017deep}. Generative adversarial networks (GAN) was employed to generate the private information from the trained deep neural network. Though differential privacy (DP) was introduced, it was not investigated if it will be effective for the new type of active inference attack. In a practical setting, Shi et al.~\cite{shi2020over} investigate the wireless systems, which utilise machine learning models for wireless signal classifier. User authentication in 5G or IoT systems is used as an example. A surrogate model, namely a functionally equivalent classifier, is built as the basis of the membership inference attacks. Another work from ~\cite{chen2020gan} studies the inference attacks against the deep generative models based on four different experiment settings, which are accessible discriminator, white-box generator, partial black-box generator and full black-box generator. Although it is not evaluated in a real-world setting, it presents a first taxonomy for membership inference attacks against GAN models. A more realistic threat model is discussed recently in ~\cite{choquette2021label}, which assumes the only access to hard labels of the trained model. It was expected this type of label-only membership inference attacks could be mounted on any machine learning-based systems. The evaluation of defence strategy suggests that at present, differentially private training and strong L2 regularization will be the most effective defence methods. Another type of inference attacks is the property inference attacks. Ganju et al.~\cite{ganju2018property} proposed to investigate the fully connected neural networks (FCNNs) against property inference attacks. Various data properties were inferred, such as the disproportion of data samples and whether the training data is noisy.\par
\indent Some works have termed the reverse-engineering work of black-box models as \textit{model extraction attacks}. One early work is by Tramer et al. ~\cite{tramer2016stealing}, who have further explored the scenario when attackers only have the query access but no prior knowledge of the models or data. The simple yet efficient attacks were devised to extract targeted ML models including logistic regression, neural networks and decision trees. Furthermore, the online services of BigML and Amazon Machine Learning were attacked, which indicated enhanced countermeasures will be needed to ensure the ML-as-a-service providers avoid model extraction attacks via exploiting common prediction APIs. Considering the architecture, optimisation process and training data as the model attributes, Oh et al.~\cite{oh2019towards} further explored the model extraction attack for the trained neural network and deep learning models. Juuti et al.~\cite{juuti2019prada} have subsequently evaluated different system factors that may relate to the success of DNN model extraction attack, and found that neither hiding hyperparameters of the target model nor reducing DNN outputs from classification probabilities to labels only could contribute to defence.\par
\indent Similar to model extraction attacks, \textit{model stealing attacks} is another major type of model-oriented attacks targeting on the confidential component, i.e., the model hyperparameters. Wang et al.~\cite{wang2018stealing} firstly considered the machine learning models including ridge regression, logistic regression, SVM and neural network as the attacking targets in a white-box setting. The threat model assumes an attacker has access to the training dataset, ML algorithm and optionally to the model parameters. The attack was further evaluated on the machine-learning-as-a-service cloud platform Amazon Machine learning and the hyperparameters value in the objective function were successfully obtained. Keeping the model architectures information and limiting the adversary access to natural seed samples are effective ways to limit the effectiveness. Thus, a generic defence approach `Protecting against DNN Model Stealing Attacks' (PRADA) to detect the model extraction attacks was proposed to detect attacks spanning several queries. The model stealing attacks targeting on the model functionality was also designed with the black-box setting in ~\cite{orekondy2019knockoff}. The knockoff models give the adversary the advantages to bypass the monetary cost and intellectual efforts. A real-world image recognition API was introduced for further evaluation and the results showed strong performance of the devised knockoff model. In ~\cite{correia2018copycat}, a copycat CNN model attack was presented by utilising random non-labeled data in two steps.\par
\indent In the work by Papernot et al.~\cite{papernot2016transferability}, two different attacks modes were investigated, which are the model extraction attacks and adversarial attacks. The experiments showed the applicability of DNN and logistic regression (LR) models to learn a substitute model for different classifiers, i.e., deep neural network, logistic regression, support vector machine, decision tree and nearest neighbors. Meanwhile, the online classifiers trained by Amazon and Google were also attacked to misclassification in a black-box setting using a logistic regression substitute model. Huang et al.~\cite{huang2021robustness} extended the \textit{adversarial attack} to the pre-trained models from TensorFlow Hub and successfully attacked all 10 representative real-world Android apps with the devised attack. The adversarial examples indicated a practical approach to attack the deep learning models while knowing their pre-trained models for TFLite models. In the work by Hitaj et al.~\cite{hitaj2019evasion}, evading the machine learning model watermarking technique is one critical step to avoid the detection and verification by legitimate owners of the stolen machine learning model, which is categorised as the model evasion attacks.\par
\begin{center}
\fontsize{8}{9}\selectfont
\begin{longtable}{|p{1.4cm}|p{2.6cm}|p{2.5cm}|p{2.5cm}|p{1.8cm}|p{2cm}|}
\caption{Summary of the surveyed works for the \textit{Model-oriented attacks} for MLBSS.}
\label{tab:model_attack}\\
\hline
\textbf{Study} & \textbf{Attack mode} & \textbf{Attack stage} & \textbf{Model} & \textbf{Affecting CIA} & \textbf{Defence}\\\hline
\endfirsthead
\multicolumn{6}{c}%
{\tablename\ \thetable\ -- \textit{Continued from previous page}} \\
\hline
\textbf{Study} & \textbf{Attack mode} & \textbf{Attack stage} & \textbf{Model} & \textbf{Affecting CIA} & \textbf{Defence}\\\hline
\endhead
\hline \multicolumn{6}{r}{\textit{Continued on next page}} \\
\endfoot
\hline
\endlastfoot
  Blanchard et al.~\cite{blanchard2017machine} & \multirowcell{5}[0ex][l]{\textit{Poisoning attacks}} & Model construction & Distributed machine learning & Model availability & None\ \\\hhline{-~*{4}{-}}
  Bhagoji et al. 2019~\cite{bhagoji2019analyzing} & & Model construction & Federated machine learning & Model availability & Krum\&Coordinate-wise median\\\hhline{-~*{4}{-}}
  Fang et al. 2020~\cite{fang2020local} & & Model construction & Federated machine learning & Model availability & ERR\&LFR \\\hhline{-~*{4}{-}}
  Kurita et al. 2020~\cite{kurita2020weight}& & Model construction & BERT & Data \& Model integrity & SHA hash checksums\&Label flip rate \\\hhline{-~*{4}{-}}
  Lorenz et al. 2021~\cite{lorenz2021backdoor} & & Model construction & DNN & Model integrity \& availability & Fine-pruning \\
  \hline  
  Ji et al. 2018~\cite{ji2018model}& \textit{Model-reuse attacks} & Model construction & DNN & Model integrity & None \\
  \hline
  Liu et al. 2017~\cite{liu2017neural} & \multirowcell{5}[0ex][l]{\textit{Trojan attacks}} & Model construction & SVM, decision tree & Model integrity & Input preprocessing, anomaly detection and re-training\\\hhline{-~*{4}{-}}
  Chen et al. 2019~\cite{chen2019deepinspect} &  & System deployment & DNN & Model integrity & DeepInspect\\\hhline{-~*{4}{-}}
  Gu et al. 2019~\cite{gu2019badnets} & & Model construction & CNN & Model integrity & None\\\hhline{-~*{4}{-}}
  Guo et al. 2019~\cite{guo2019tabor} &  & System deployment & DNN & Model integrity & TABOR\\\hhline{-~*{4}{-}}
  Costales et al. 2020~\cite{costales2020live} &  & System deployment & CNN & Model integrity & None \\
  \hline
  Ateniese et al. 2015~\cite{ateniese2015hacking} & \multirowcell{8}[2ex][l]{\textit{Model inference attacks}}& System deployment & SVM, Hidden Markov models& Data confidentiality& None \\\hhline{-~*{4}{-}}
  Hitaj et al. 2017~\cite{hitaj2017deep} & & Model construction & Distributed deep learning & Data confidentiality & None \\\hhline{-~*{4}{-}}
  Shokri et al. 2017~\cite{shokri2017membership} & & System deployment& CNN  & Data confidentiality & Regularisation \\\hhline{-~*{4}{-}}
  Ganju et al. 2018~\cite{ganju2018property}& & System deployment& FCNNs  & Data confidentiality & None \\\hhline{-~*{4}{-}}
  Salem et al. 2019~\cite{salem2019ml}& & System deployment & CNN & Data confidentiality & Dropout \& model stacking\\\hhline{-~*{4}{-}}
  Shi et al. 2020~\cite{shi2020over}& & System deployment & DNN & Data confidentiality & None \\\hhline{-~*{4}{-}}
  Chen et al. 2020~\cite{chen2020gan}& & Model construction & GAN & Data confidentiality & Differential privacy \\\hhline{-~*{4}{-}}
  Choquette et al. 2021~\cite{choquette2021label}& & Model construction & DNN & Data confidentiality & Differential privacy, strong l2 regularization\\
  \hline
  Tramer et al. 2016~\cite{tramer2016stealing}& \multirowcell{6}[2ex][l]{\textit{Model extraction attacks}} & System deployment & Logistic regression, neural network and decision tree & Model confidentiality & None\\\hhline{-~*{4}{-}}
  Oh et al. 2019~\cite{oh2019towards}& & System deployment & CNN & Model confidentiality & None\\\hhline{-~*{4}{-}}
  Juuti et al. 2019~\cite{juuti2019prada}& & System deployment & DNN & Model confidentiality & PRADA \\\hhline{-~*{4}{-}}
  Correia et al. 2018~\cite{correia2018copycat}& & System deployment & CNN & Model confidentiality & None \\\hhline{-~*{4}{-}}
  Krishna et al. 2020~\cite{krishna2020thieves} & & System deployment & BERT & Model confidentiality & Watermarking \\\hhline{-~*{4}{-}}
  Jagielski et al. 2020~\cite{jagielski2020high} & & System deployment & DNN & Model confidentiality & None\\
  \hline
  Wang et al. 2018~\cite{wang2018stealing}& \multirowcell{2}[0ex][l]{\textit{Model stealing attacks}} & System deployment & Ridge regression, logistic regression, SVM and NN & Model confidentiality & None \\\hhline{-~*{4}{-}}
  Orekondy et al. 2019~\cite{orekondy2019knockoff}& & System deployment & CNN & Model confidentiality & None \\\hline
  Papernot et al. 2016~\cite{papernot2016transferability}& Adversarial \& Model extraction attacks & System deployment & DNN, LR, SVM and decision trees & Model confidentiality; data integrity & None\\\hline
  Huang et al. 2021~\cite{huang2021robustness}& Adversarial attacks & System deployment & DNN & Model integrity & None\\\hline
  Hitaj et al. 2019~\cite{hitaj2019evasion}& Evasion attacks (Ensemble attack, detector attack)& System deployment & CNN & Model confidentiality & None \\\hline
\end{longtable}
\end{center}
\subsection{System-oriented attacks}\label{subsec:system_attacks}
Till now, the analysis of security threats against MLBSS centers on the data, algorithmic and model aspects. As we have discussed in previous sections of `data-oriented attacks', `model-oriented attacks', in this section, the focus is on the vulnerabilities from the constructional and functional components carried with the system. Since MLBSS will harness more dedicated and dynamics components in terms of hardware and software to fulfill the dedicated functionalities, these components are tightly connected to achieve the legitimate operations in a valid and sustainable way, i.e., concerning the increasing computation and storage requirements, and the run-time environment. The traditional security threats have become much more complicated in the MLBSS, in which new attack surface can be from the middleware, the globalised supply chain, the runtime engine and so on. However, the system-oriented security threats often emerge through the interplay of multiple systems while overseen until they are actively exploited. While the attackers may leverage various attacking surfaces, including hardware, in this section, we will discuss recently identified system-oriented attacks.\par
\textbf{Review \& discussion:}\par
In this work, we firstly measure the hardware-related system-oriented attacks with respect to the goals and capabilities that alter the model and data information via manipulating the hardware-related software and channels, i.e., the middleware. One instrumental asset that largely contributes to the success of MLBSS is the advanced computational infrastructure, i.e., the specialised hardware accelerators. Particularly, recent advancement of machine learning techniques has been well positioned with a large demand of computing power to handle the accumulated big dataset and complicated calculations for deep neural networks. It is now a hot research topic aiming at better leveraging not only the software, but also the hardware, in terms of the design logic to fulfill the proposed functionalities. For example, specialised hardware accelerators are one type of the hardware that upholds the hardware requirements for such resource-demanding applications. Accordingly, system-on-Chip (SOC) design has become a hot topic to ensure the model performance and improve the energy efficiency, which has exposed further stealthy attack surfaces, i.e., implanting the hardware trojan in the Graphical Processing Units (GPUs) and Tensor Processing Units (TPUs). Such an attacker can leverage the hardware attack vectors to achieve the goals of model corruption, backdoor insertion, model extraction, spoofing, information extraction and sybil attack~\cite{xu2021security}. Generally, the traditional and popular process for system designers to solve the hardware-software co-design problem is to integrate specialised hardware accelerators from external hardware designers. This results in a significant concern of trust, since outsourced accelerators could have malicious modiﬁcations, also known as hardware Trojan. Several works have extensively discussed this matter covering the traditional ASIC and FPGA architectures ~\cite{ye2018hardware,yang2020hardware,xue2020ten,hu2021practical,liu2020sequence,collantes2020safetpu,liu2017fault,venceslai2020fault}. Given the continuous development and progress of hardware-software co-design for MLBSS, the hardware-oriented approaches attempting to steal or corrupt the hardware thus impacting the MLBSS have seen its severity.\par
\indent Furthermore, we specifically focus on the hardware-related software attack vectors. One extraction attack on the unencrypted PCIe traffic is summarised in Table.~\ref{tab:system_attack}~\cite{zhu2021hermes}. The unencrypted PCIe traffic is a novel side-channel attack surface~\cite{xu2021security,zhu2021hermes}. Different from general model extraction methods, the hardware model stealing attack assumes that the adversary has physical access to GPU devices, particularly to the PCIe bus. A novel Hermes Attack was proposed to fully reconstruct the whole DNN models in a black-box setting. The last group of hardware-related attacks is discussed in ~\cite{xu2021security,mittal2021survey}, which is side-channel attacks. The goals of side-channel attacks can be various, including model extraction from PCIe~\cite{zhu2021hermes} and GPU~\cite{naghibijouybari2018rendered,wei2020leaky}.\par

\begin{table}
\fontsize{8}{9}\selectfont
\centering
\caption{Summary of the surveyed works for the \textit{System-related attacks} for MLBSS.}
\label{tab:system_attack}
\begin{tabular}{|p{1.5cm}|p{2.5cm}|p{2.5cm}|p{2.5cm}|p{1.8cm}|p{2cm}|}
  \hline \textbf{Study} & \textbf{Attack mode} & \textbf{Attack stage} & \textbf{Model} & \textbf{Affecting CIA} & \textbf{Defence}\\\hline
  Song et al. 2017~\cite{song2017machine} & User privacy stealing & System deployment & CNN, Residual networks, SVM \& LR & Data confidentiality & Data obfuscation \\\hline
  Liu et al. 2019~\cite{liu2019system}& System-level DNN attacks & Model construction & DNN & Model integrity \& confidentiality & None \\\hline
  Kim et al. 2020~\cite{kim2020impact}& Platform attacks & System deployment & CNN & Model availability & Data-space randomization, hardware access controls \\\hline
  Costa et al. 2021~\cite{costa2021covert}& Covert channels attacks & Model construction & Federated machine learning & Model integrity & None \\\hline
  Xu et al. 2021~\cite{xu2021security}& Trojan, side-channel \& fault-injection attacks& Model construction & Neural network & Model integrity\&availability & Yes\\\hline
  Mittal et al. 2021~\cite{mittal2021survey}& Trojan, side-channel \& fault-injection attacks & Model construction & DNN & Model integrity \& availability & Yes \\\hline
  Zhu et al. 2021~\cite{zhu2021hermes}& Side-channel attacks & System deployment & DNN & Model confidentiality & Encryption, obfuscation, adding noise\&offload tasks to CPU \\\hline
\end{tabular}
\end{table}

\indent In comparison to the algorithmic attacks for machine learning-based software systems, another observation is that the system-oriented attacks can also be easily integrated with traditional attack vectors and approaches like malware injection and user privacy stealing. Since machine learning has become a commodity, Song et al.~\cite{song2017machine} studied the scenario that machine learning code was provided by a malicious service provider. It was identified that, no matter the adversary is on white- or black-box mode, the provided machine learning code can be the attacking surface that ``memorise'' the sensitive information from the user in the final model. The model was still proved to be accurate and meet the functional requirements. However, it leaves the exploitable point for the adversary to obtain the privacy. It would be relatively difficult for the success of manual inspection of the source code. One suggested manner from ~\cite{song2017machine} is to turn the least significant bits attacks against itself. In ~\cite{liu2019system}, an overview of the system-level attacks is discussed, particularly focusing on malware injections. An assumption was made that the adversary may have the exclusive access to the models, as well as the supporting hardware and software components in the corresponding execution engine. Following, Kim et al.~\cite{kim2020impact} further elaborated the attacks towards AI platforms. A memory safety vulnerability in a third party library was investigated which has caused the targeted misclassification for the machine learning system. The open-source toolkit, namely Microsoft CNTK, was leveraged and the exploit relied on the identified vulnerability of CVE-2018-5268. Recently, Costa et al.~\cite{costa2021covert} utilised the federated machine learning system to introduce a new covert channels attack. The stealth communications between learning agents were established by violating the isolated training protocol, which could lead to the user privacy stealing and model manipulation.\par
\subsection{Summary}
\indent In this section, we have collectively surveyed the security threats for machine learning-based software systems with a holistic system view, which is illustrated as well in Fig.~\ref{fig:mlbss_taxonomy} covering data- (15 papers), model- (30 papers), and system-oriented (7 papers) attacks. It is noted that, current the attacks and defences for MLBSS are more and more similar as the cat and mouse games in traditional system security area. While the collection of the papers can be more due to the popularity of this topic, we anticipate that the collective discussion in this section covers the major distinguishable artefacts of MLBSS. The attacks modes are subsequently categorised in terms of the direct or most related artefact.\par
\indent Following the studies from ~\cite{piessens2002taxonomy,mcgraw2020top}, Table.~\ref{table:traditional_taxonomy} is a summarised two-level hierarchy taxonomy of vulnerabilities causes for general internet software systems~\cite{piessens2002taxonomy}. While the classical taxonomy is not a static item, it serves as a general reference to classify the software security threats in terms of software development practice, which are secure input and output handling, race condition, faulty use of API, memory safety, improper use case handling, improper exception handling, resource leaks, and pre-processing input strings~\cite{piessens2002taxonomy,weicomprehensive}. Thus, the risks of MLBSS depicted in Table.~\ref{tab:top_ten_risks}~\cite{mcgraw2020top} are exclusively covered and extended in the discussion of the literature in this section with a taxonomy summary of Fig.~\ref{fig:mlbss_taxonomy}. Although ~\cite{mcgraw2020top} presented the analysis mostly from a general term, we have substantially extended the scope and distill the available details from relevant literature to justify the emerging and novel attack modes.\par
\indent In the following sections, we will continue the investigation with security threats with the traditional software engineering practices for MLBSS as many challenges have been proposed. With the discussion from the perspective of affecting developing stage, it is anticipated to understand the state-of-the-practice of secure development for MLBSS. A major part of the discussion will subsequently focus on the secure development practice, introducing the main strategies for MLBSS development lifecycle.\par
\begin{table}
\fontsize{8}{9}\selectfont
\caption{A Two-level hierarchy taxonomy of vulnerabilities causes for traditional internet software systems~\cite{piessens2002taxonomy}}
\begin{tabular}{|p{5cm}|p{8cm}|}\hline
\textbf{Phases} & \textbf{Vulnerabilities Causes} \\\hline
\multirow{3}{*}{Analysis phase}& No risk analysis/ No security policy \\\cline{2-2}
 & Biased risk analysis \\\cline{2-2}
 & Unanticipated risks \\\hline
\multirow{4}{*}{Design phase}& Relying on non-secure abstractions \\\cline{2-2}
 & Security/Convenience tradeoff \\\cline{2-2}
 & No logging\\\cline{2-2}
 & Design does not capture all risks\\\hline
\multirow{6}{*}{Implementation phase}& Insufficiently defensive input checking \\\cline{2-2}
 & Non-atomic check and use\\\cline{2-2}
 & Access validation errors\\\cline{2-2}
 & Incorrect crypto primitive implementation\\\cline{2-2}
 & Insecure handling of exceptional conditions\\\cline{2-2}
 & Bugs in security logic\\\hline
\multirow{3}{*}{Deployment phase}& Reuse in more hostile environments \\\cline{2-2}
 & Complex or unnecessary configuration\\\cline{2-2}
 & Insecure defaults\\\hline
\multirow{2}{*}{Maintenance phase}& Feature interaction\\\cline{2-2}
 & Insecure fallback\\\hline
\end{tabular}
\label{table:traditional_taxonomy}
\end{table}
\begin{center}
\begin{table}
\fontsize{8}{9}\selectfont
\centering
\caption{List of Top Ten Risks of Machine learning-based Systems~\cite{mcgraw2020top}}
\label{tab:top_ten_risks}
\begin{tabular}{|p{4.5cm}|p{4.5cm}|}
  \hline Adversarial examples & Data poisoning\\\hline
  Online system manipulation & Transfer learning attack\\\hline
  Data confidentiality & Data trustworthiness\\\hline
  Reproducibility & Overfitting\\\hline
  Encoding integrity & Output integrity\\\hline
\end{tabular}
\end{table}
\end{center}

\section{Secure development practices for MLBSS}
\label{sec:security_practice}
To understand the security threats for MLBSS, a thorough review via the traditional confidentiality, integrity and availability (CIA) model has been implemented with respect to the attacks types towards the dominant artefacts in MLBSS in Section.~\ref{sec:taxonomy} (see Table.~\ref{tab:data_attack}, ~\ref{tab:model_attack}, and ~\ref{tab:system_attack}). Meanwhile, the attacking stages are discussed considering the MLBSS system development lifecycle from problem definition, data management, model construction, system deployment to system maintenance. It is observed that current attacks have a major focus on data management and model construction stages. It has now aimed for a more stealthier approach and a more critical outcome with either additional access to the side channels or advanced algorithms to learning outcomes. Human factors remain remain the fundamental element of system security~\cite{assal2019think}.\par
\indent Thus, in this work, we have further investigated how current software engineering practices can support the system security of MLBSS by mitigating the aforementioned various attack types and outcomes. Since the integration of security into the system lifecycle can serve an efficient insertion of security processes into the existing system lifecycle~\cite{davis2005secure,radack2009system,aljawarneh2017cloud}, we briefly describe the current software engineering practices for MLBSS in general. A main discussion will follow the MLBSS lifecycle, particularly from the perspective of the security practices.\par
\subsection{Software engineering practices for MLBSS}
Whilst the machine learning components have extensively ongoing development and become an integral component in larger software systems, the wide deployment of MLBSS has led the research trends to adopt various software engineering practices. Extending the development process with the essential components from ~\cite{sculley2015hidden}, in Fig.~\ref{fig:mlbss_component}, the machine learning model development only represents a tiny fraction of whole system. It is undoubted that the intertwining process of machine learning and software engineering has opened many challenges. A recently reported industrial case SAP~\cite{rahman2019machine} reports the challenges in software engineering.\par
\indent For traditional software development, the Guide to the Software Engineering Body of Knowledge (SWEBOK)~\cite{bourque2014guide} provides a comprehensive umbrella to cover the software engineering topics, including \textit{software requirement}, \textit{software design}, \textit{software construction}, \textit{software maintenance}, \textit{software engineering models and methods}, \textit{software engineering process}, \textit{software quality}, \textit{software testing}, \textit{software engineering management}, \textit{software configuration management} and \textit{software engineering professional practice}. To understand the difference and challenges, SWEBOK is introduced as the taxonomy for SE practices classification in ~\cite{wan2019does,kumeno2019sofware,amershi2019software,serban2020adoption,martinez2021software}. A conclusion is collated that, the challenges presented in the development of MLBSS are specific and can not be well managed within the SWEBOK topics, especially in terms of the technical and data-related challenges. This, in turn, calls for adjusting the practices in the context of system domain and technical issues.\par

\begin{figure}[htbp]
  \centering
  \includegraphics[width=0.8\linewidth,keepaspectratio]{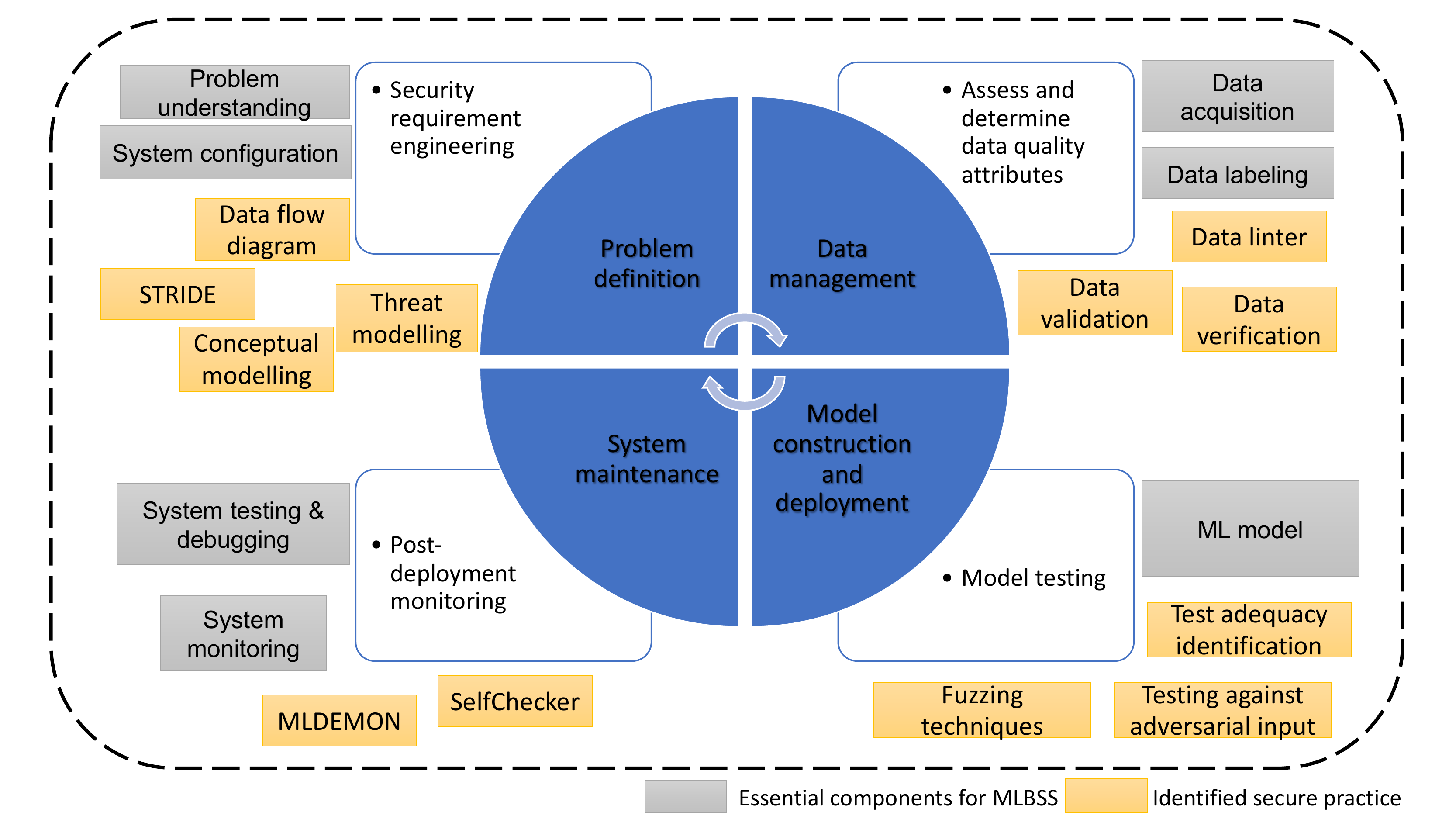}
  \caption{The essential components~\cite{sculley2015hidden} and identified secure practices for MLBSS}
  \label{fig:mlbss_component}
\end{figure}

\indent Currently, the research and development of the state-of-the-practice for MLBSS are mostly focused on two topics, which are technical debt and quality assurance. Generally, quality assurance defines `\textit{a planned and systematic pattern of all actions necessary to provide adequate confidence that the item or project conforms to established technical requirements}'~\cite{buckley1984software}. For technical debt, at its broadest, it was defined as `any side of the current system that is considered sub-optimal from a technical perspective'~\cite{ktata2010designing,tom2013exploration}. While the discussion of technical debt is related to the compromised decisions for the poor software development~\cite{liu2020using}, the quality assurance dedicates to system quality offline and in production attached with several distinct aspects, i.e., model quality, safety and fairness analysis~\cite{kastner2020teaching}.\par
\indent Technical debt, studied in ~\cite{tom2013exploration} for classical software systems, has recently been reviewed in MLBSS in order to identify the specific patterns~\cite{sculley2015hidden,liu2020using,tang2021empirical,bogner2021characterizing}. Additional to typical technical debt at code level, the machine learning-based systems represent extra trade-offs to be overcome for practical practices for the long term development, deployment and maintenance at the system level. The data dependency debt was extensively discussed as a key contributor for MLBSS since it is one of the integral components in MLBSS and is generally more difficult to detect as a result of its unstable and underutilised characteristics~\cite{sculley2015hidden}. Other technical debts, i.e., abstraction debt, configuration debt, data testing debt, reproducibility debt, process management debt and cultural debt, may also be accrued and demands ongoing collaborative efforts. Following in ~\cite{bogner2021characterizing}, 21 primary studies were included to identify the actual technical debts that have been investigated in recent works, in which four novel emerging debts of data, model, configuration and ethics were elaborated. Taking further actions towards measuring the technical debt and targeting to pay it off, a recent work by Liu et al. has investigated the deep learning frameworks which is an essential component for both researchers and industry developers~\cite{liu2020using}. The analysis of seven most popular open-source deep learning frameworks illustrates a demanding understanding and countermeasures for the design debt, defect debt, documentation debt, test debt, test debt, requirement debt, compatibility debt and algorithm debt. In ~\cite{tang2021empirical}, refactoring was empirically studied against MLBSS, particularly to answer the machine learning-specific refactoring. It is found that duplicate code elimination in configuration and model code is the top refactoring for code duplication debt in MLBSS, while more refactoring will be required to technical debt in MLBSS.\par
\indent Another main topic is about quality assurance. Quality assurance has been well established for traditional software development in terms of proper software quality in conformance with standards and procedures~\cite{buckley1984software}. Ma et al.~\cite{ma2018secure} is the first work that attempts to answer the question of common challenges relevant to quality assurance of deep learning. Particularly, a collection of existing literature was evaluated indicating that there still lacks a thorough understanding of the occurrence of deep learning bugs. Thus, a Secure Deep Learning Engineering Life Cycle (SDLE) was introduced to support the future quality assurance of development. Recently, Hamada et al.~\cite{hamada2020guidelines} further investigated the quality assurance for machine learning-based systems by reporting the initial version of guidelines, which was an extended outcome of QA4AI Consortium held in Japan consisting 39 researchers and practitioners and three different organisations. The common core concepts of the quality assurance for MLBSS were summarised in terms of the axes of quality evaluation. Considering two different software development styles, i.e., deductive style and inductive style, five aspects of quality evaluation for MLBSS, including data integrity, model robustness, system quality, process agility and customer expectation, were extracted.\par
\indent Including the discussed technical debt and quality assurance for MLBSS, the development of MLBSS remains challenging to define a comprehensive and consolidated engineering protocol, which will ensure a robust and secure development, deployment and maintenance process. Some works have attempted to explore the distinct components from system level. Humbatova et al. looked into the general artefacts, i.e., \textit{model}, \textit{tensor\&input}, \textit{training}, \textit{GPU usage} and \textit{API usage}, which may be failed and exploited in deep learning systems from the developer's points of view~\cite{humbatova2020taxonomy}. Lewis et al. alternatively took the mismatch system elements that may occur during the end-to-end development by different stakeholders into the consideration, in which the trained model, raw data, training data and the environments were identified~\cite{lewis2021characterizing}. However, more efforts are still needed to support the robust and secure engineering of MLBSS. In following subsection, we will reviewed the current security practices for MLBSS.\par
\subsection{Current secure development practices for MLBSS}
The goal of system security for MLBSS is to ensure the software systems function in a correct way continuously in the deployed environment~\cite{mcgraw2004software}. It engages the software engineering practices by distilling the knowledge of security threat and risk analysis in the software development lifecycle~\cite{lee2002integrating}, such as early on for the security requirements engineering~\cite{haley2008security,mellado2010systematic}. The practices of security engineering would need to be applied to various software artefacts to understand their adoption and affects accordingly within the system context.\par
\indent One recent work by McGraw et al.~\cite{mcgraw2019security} demonstrates that the security risks for MLBSS exhibit a much more complicated scenario for all practitioners. A thorough understanding of the artefacts, along with their inherent relationship between each other and the external interactions with the process, is the main driving force towards securing machine learning-based software systems. McGraw et al. has subsequently focused on the architectural risk analysis for machine learning security, which firstly initiated the work of ~\cite{mcgraw2020architectural} and further elaborated the identified top ten risks in~\cite{mcgraw2020top}. While the analysis output can serve as a risk analysis framework for MLBSS, it lacks an in-depth consideration to design, implement and field the machine learning-based systems.\par
\indent Noting that a brief discussion of security risks analysis for data, model (development and runtime), input (runtime) from ~\cite{mcgraw2019security}, we have further categorised and curated the identified attacks targeting on data, model, and systems through Section~\ref{sec:taxonomy}. This serves a refined representation of these specific attacks from the artefacts perspective, which could be better leveraged for MLBSS secure development practices from the software lifecycle. As illustrated in Fig.~\ref{fig:mlbss_component}, current secure development practices for MLBSS also share same conceptual maps to some extent while there remains much more work unexplored to bridge the novel attack modes and challenges. Currently, significant researches targeting secure development practices for MLBSS are under development. Accordingly, we review the work focusing on each development phase, including \textit{security requirement engineering} for problem definition, \textit{data verification and sanitization} for data management, \textit{machine learning model and dependency testing} for model construction and deployment, and \textit{post-deployment monitoring} for system maintenance. We will also include the discussion and vision of these work concerning the mitigation of categorised attacks from the artefacts perspective.\par

\subsubsection{\textbf{Problem definition}}\label{subsec:problrm_definition}\par
The different application trends in the software engineering industry demand different understanding and practice, which propels the development process to be in dynamically changes regarding the daily activities and the practitioners' roles evolution. It is critical to properly define the problem in the context of application. In the stage of \textit{problem definition}, designers and related stakeholders are gathered together to determine the features concerning the functional and non-functional requirements, particular when the systems require machine learning driven modules as the core. This brings the \textit{security requirement engineering} a critical software development process to general information systems~\cite{mellado2010systematic}.\par
\indent Whilst taking security requirement engineering early on in the modern software development process, the inherent uncertainty of machine learning driven modules is one of the main constraints that impose critical challenges and substantial changes to the existing secure development practice~\cite{amershi2019software}. Recently, several studies have aimed to address the requirement engineering for machine learning, particularly from the security requirement engineering aspect~\cite{horkoff2019non,wilhjelm2020threat,gauerhof2020assuring,nalchigar2021modeling,yang2021exploiting,heyn2021requirement}.\par
\indent One early study ~\cite{horkoff2019non} initiated the discussion of quality-related non-functional requirement for machine learning-based systems, for which significant challenges and gaps were identified for the qualities of machine learning-based systems. It was the first work related to the qualities for MLBSS such as security, privacy and transparency and so on. Wilhjelm et al.~\cite{wilhjelm2020threat} firstly extended the security concerns for MLBSS with a holistic paradigm to include the specific and traditional threats. Particularly, Wilhjelm et al. has applied the traditional threat modelling method together with the identified attack libraries to understand the security of MLBSS at the requirements phase~\cite{wilhjelm2020threat}. Data flow diagrams and STRIDE methods were utilised as the main methods for security requirements elicitation in practices. STRIDE is one threats classification methodology based on the Spoofing, Tampering, Repudiation, Information Disclosure, Denial of Service, and Elevation of Privilege threat categories. Conceptual modelling is introduced in ~\cite{nalchigar2016conceptual,nalchigar2017conceptual} to provide multi modelling views for systematic analytical requirements elicitation, which are the business view, analytics design view and data preparation view. It is subsequently employed in the health sector~\cite{nalchigar2021modeling}, though the details concerning the quality attributes or non-functional requirements for users are not thoroughly discussed in terms of the gained advantages via the goal-oriented requirement for machine learning (GR4ML) framework. \par
\indent Another work from ~\cite{gauerhof2020assuring} employed a different process defined in ~\cite{ashmore2021assuring} to the machine learnt models (MLMs) in self-driving cars. The safety requirements for MLMs were analysed to drive the assurance activities for data management and model construction phases. It is noted that a rigorous and accurate understanding and definition of the problem could well serve as a mitigation method thwart the success of \textit{data-} and \textit{model-oriented} attacks (see section. ~\ref{subsec:data_attacks},~\ref{subsec:model_attacks}). It is demonstrated with a specific task of pedestrian detection at crossings by considering the vehicle-level safety requirements. The MLMs safety requirements for data management phase were subsequently elicited in line with the assurance desiderata, which are relevant, complete, accurate and balanced. However, the details of model construction phase were missing. While there is no other existing work to assure MLMs as well as the released safety standards such as ISO26262, Gauerhof et al.~\cite{gauerhof2020assuring} has demonstrated the importance and applicability of designing MLMs safety requirements for the safety assurance activities in data management and model construction phases. Yang et al. considered the specific safety-critical system for machine learning deployment, i.e., the aerospace domain~\cite{yang2021exploiting}. The restricted natural language requirements modelling process was evaluated with the implementation of an industrial case Autonomous Guidance, Navigation and Control (AGNC). A most recent research from ~\cite{heyn2021requirement} has illustrated that to guarantee the machine learning-based systems behaviour with a well-defined system attributes such as safety, major problem areas including contextual definition and requirements, data attributes and requirements, performance definition and monitoring, and human factors, will need substantial efforts from the requirement and systems engineering research.\par

\subsubsection{\textbf{Data management}}\label{subsec:data_management}\par
Data persistence is important to the entire software systems. For traditional systems, the privacy of the digital information such as user data and system configuration in modern systems is critical and demands thorough and effective security strategies for the entire lifecycle~\cite{yang2020data}. While it is a general requirement that the data security covers the hardware and general software logic aspects~\cite{ibm2021security}, the interaction with machine learning models requires exclusive consideration of data and its probability distribution representation~\cite{heyn2021requirement}. It is due to the uncertainty of the learnt machine learning models, which brings the data quality~\cite{batini2007framework,heinrich2018requirements} as one important aspect during data management stage to ensure the development of MLBSS. Regarding data as the object to be utilised in both design time and runtime, how to assess and determine the data quality attributes in the context of machine learning-based systems carry the significant value for the data-oriented attacks towards a robust and reliable model and sytem~\cite{willers2020safety,song2021auto,sambasivan2021everyone}.\par
\indent Overall, there have been some state-of-the-art practices targeting on the data management tasks against data-oriented attacks, such as the enrichment of dedicated data metrics and the study of data and model correlation for the data poisoning attacks~\cite{zhao2020distributed,nurminen2019software,foidl2019risk}, a data schematic for adversarial attacks~\cite{polyzotis2017data,zhang2020towards,lwakatare2021experiences} and data practices of lint and strictly controlled labelling process for backdoor attacks~\cite{hynes2017data,prendki2018curse}. An initial attempt to support the data management task for machine learning-based system in a comprehensive manner is from~\cite{polyzotis2017data}. A high-level schematic compromising training/serving data and model for a production machine learning pipeline was discussed, which further elaborated four key challenges for data management task, which are understanding, validation, cleaning and enrichment. Data validation is considered as one prominent process to guarantee data quality for MLBSS~\cite{lwakatare2021experiences}. While it has potential thwart the data poisoning and adversarial attacks, it remains unknow for the data backdoor attacks. Hynes et al.~\cite{hynes2017data} explored the common data practices for developing machine learning models by introducing the concept of lint for data preparation and feature engineering. Data linter was designed and implemented as a tool to support the analysis of training data for miscoding, outliers and packaging errors, thus the targeting model quality could be ensured and improved to minimise the chances of backdoor attacks. Prendki et al.~\cite{prendki2018curse} investigated the data labelling process to be performed in a strictly controlled process by discussing three common practices for reducing the data annotation costs in the lifecycle. While the data quality can not be ensured, i.e., excluding faulty data from different sources, Nurminen et al.~\cite{nurminen2019software} studies the correlation between the complex data and different machine learning models to support the development of robust and reliable MLBSS. Another way to mitigate the low data quality issue is a conceptual risk-based data validation approach~\cite{foidl2019risk}.\par
\indent In spite of ensuring the data quality is critical, there are limited research works targeting understanding the violation of CIA model in data management stage. Other contemporary methods to improve the robustness and reliability of machine learning-based systems include introducing extra metrics for the input data~\cite{zhang2020towards} and devising a novel verification scheme in the production environment~\cite{zhao2020distributed}. While the work by Zhao et al.~\cite{zhao2020distributed} developed a novel data verification scheme for distributed machine learning to ensure the integrity of training data, Zhang et al.~\cite{zhang2020towards} considered various uncertainty metrics for the data, particularly for the adversarial examples and benign examples. The idea is to characterise the uncertainty patterns of data for testing. One inspiring finding is that the patterns of the data generated by existing methods follow common patterns. In light of this, a new input data generation method was devised to generate uncommon adversarial examples, which experiments indicated an increasing successful rate against existing defence techniques. \par

\subsubsection{\textbf{Model construction and System deployment}}\label{subsec:model_system}\par
In comparison with traditional software systems, the MLBSS brings the learning advantages as well as the inherent uncertainty, thereby widening the potential attack surface. In Sec.~\ref{sec:taxonomy}, many attackers have demonstrated an advanced leverage of the statistical uncertainty of the trained models from various points, i.e., feature vectors and model training algorithms. Also, the system deployment have introduced novel frameworks and libraries from different sources. From one aspect, the increasing attack surfaces for MLBSS impose specific attacks for existing detection and defence techniques. From another aspect, this brings the challenges to traditional software testing strategy, which is designed as one primary instrument to evaluate system from the safety assurance perspective.\par
\indent To fulfill the traditional software testing, general practices have to identify testing principles such as test adequacy in compliance with established standards, such as ISO26262~\cite{iso201126262} for road vehicles domain. Adopting the testing practice to secure the machine learning-based systems development, particular for deep learning, has been an emerging and challenging topic for model-oriented attacks~\cite{braiek2020testing} (see section~\ref{subsec:model_attacks}). \par
\indent One studied criteria is related to the testing adequacy. DeepXplore is one of the early works that attempts to automatically generate test inputs for deep learning models to expose erroneous behaviors in an efficient way~\cite{pei2017deepxplore}. Following, different works based on traditional testing adequacy coverage methods are proposed and extensively evaluated, such as the mutation testing~\cite{ma2018deepmutation}, combinatorial testing technique~\cite{ma2019deepct}, multi-granularity testing criteria~\cite{ma2018deepgauge}. In ~\cite{kim2019guiding}, a novel test adequacy criterion for testing called Surprise Adequacy for Deep Learning Systems (SADL) is proposed. To capture the importance of deep learning-based system in a layer-wise functional understanding, ~\cite{gerasimou2020importance} has recently proposed an importance-driven test adequacy criterion. Furthermore, it was evaluated against the adversarial generation techniques for the model adversarial attacks. The other criteria could be obtained with the fault coverage for MLBSS. It could be achieved by applying dual modular redundancy to check the fault coverage in the entire DNN model~\cite{li2019d2nn,li2020deepdyve}. In~\cite{li2019d2nn}, the framework of D2NN is proposed for checking the result inconsistency between the original and duplicated models. In another work of~\cite{li2020deepdyve}, a dynamic verification method named DeepDyve is experimented with sufficient fault coverage yet little computational overhead for DNN-based classification systems. Both works have demonstrated the verification of fault tolerant capability in the DNN models could secure the model from external attacks. \par
\indent In addition to the general test principles coverage, one specific challenge for MLBSS is the robustness for adversarial input. Thus, another group of practices targets on testing MLBSS against adversarial attacks (see section~\ref{subsec:model_attacks}). Wicker et al.~\cite{wicker2018feature} initiated the adversarial examples crafting as a two-player turn-based stochastic game to test the deep neural network. Subsequently, Ma et al.~\cite{ma2019nic} has identified two common exploitation channels of the provenance channel and the activation value distribution channel in details. A novel invariant based detection technique against the adversarial samples was then proposed. The traditional coverage-guided fuzzing technique was combined with property-based testing to extend the testing coverage on adversarial samples for neural network~\cite{odena2019tensorfuzz}, while in ~\cite{xie2019deephunter} DeepHunter based on same fuzzing technique but different mutation strategies was evaluated against general-purpose DNNs.\par
\indent The third topic related to the MLBSS is the deep learning libraries, which has recently been extensively studied leading to the model-reuse attacks (see section~\ref{subsec:model_attacks}) and system-oriented attacks (see section~\ref{subsec:system_attacks}). Since the libraries are the basic component during the model construction as well as the system deployment, it will cause more critical bugs and be easier to be exploited in comparison to the model and data, such as the TensorFlow~\cite{zhang2018empirical}. Given the large amounts of different machine learning frameworks and platforms, particularly for deep learning, Guo et al.~\cite{guo2019empirical} studied the performance metrics for different frameworks. For the development process, this includes the accuracy, adversarial robustness of trained models based on four representative frameworks, which are TensorFlow 1.12.0, PyTorch-0.4.1, CNTK-2.6 and MXNet-1.4.0. Additionally, Guo et al. also studied performance after model migration and quantization for the system deployment process, such as the migrated model for mobile and web platform. With the observed performance decline for the two different processes, the concern and attention for deep learning software community are raised to address alleviate the deviation amid frameworks and platforms for a universal deep learning solutions. Pham et al. has implemented CRADLE, a cross-backend validation to detect and localise bugs in deep learning libraries~\cite{pham2019cradle}. The effectiveness of CRADLE in detecting bugs and unique inconsistencies for TensorFlow, CNTK and theano has called for extra attention for testing deep learning implementations. Similarly, oracle approximations as a common practice was adopted to represent numerical errors in deep learning libraries and was studied in four popular libraries, which are Tensorflow, theano, PyTorch and Keras~\cite{nejadgholi2019study}. LEMON, a deep learning library testing via guided mutation based on novel heuristic strategy is recently proposed by Wang et al.~\cite{wang2020deep} to library testing. It brought the attention to the 24 novel bugs which were identified in 20 published versions of 4 representative deep learning libraries such as TensorFlow, Theano, CNTK, MXNet.\par
\subsubsection{\textbf{System maintenance}}\label{subsec:system_maintenance}\par
For system maintenance, model and system monitoring is critical to the consistent assurance of MLBSS for security, reliability and efficiency. The output must be intensively monitored as a deviated output from a machine learning-based software system, which could not be fully pledged to avoid uncertainty, will cause catastrophic consequences. While the security engineering practices for MLBSS is still in its infancy, several works have attempted to monitor the model update for post-deployment.\par
\indent In ~\cite{ginart2021mldemon}, Ginart et al. investigated a much complex scenario for post-deployment monitoring, which lacks sufficient sources for labeled data to estimate the model's on-time performance. Alternatively, the MLDEMON, for ML DEployment MONitoring, was validated on eight realistic data streams to improve the system deployment in terms of reliability and efficiency. Xiao et al.~\cite{xiao2021self} focuses on the DNNs monitoring. In this work, SelfChecker is devised to monitor the outputs and triggers an alarm when inconsistency is detected in the internal layers.\par

\section{Challenges and future directions for security engineering of MLBSS}
\label{sec:challenges_futuredirections}
\begin{figure}[htbp]
  \centering
  \includegraphics[width=0.7\linewidth,keepaspectratio]{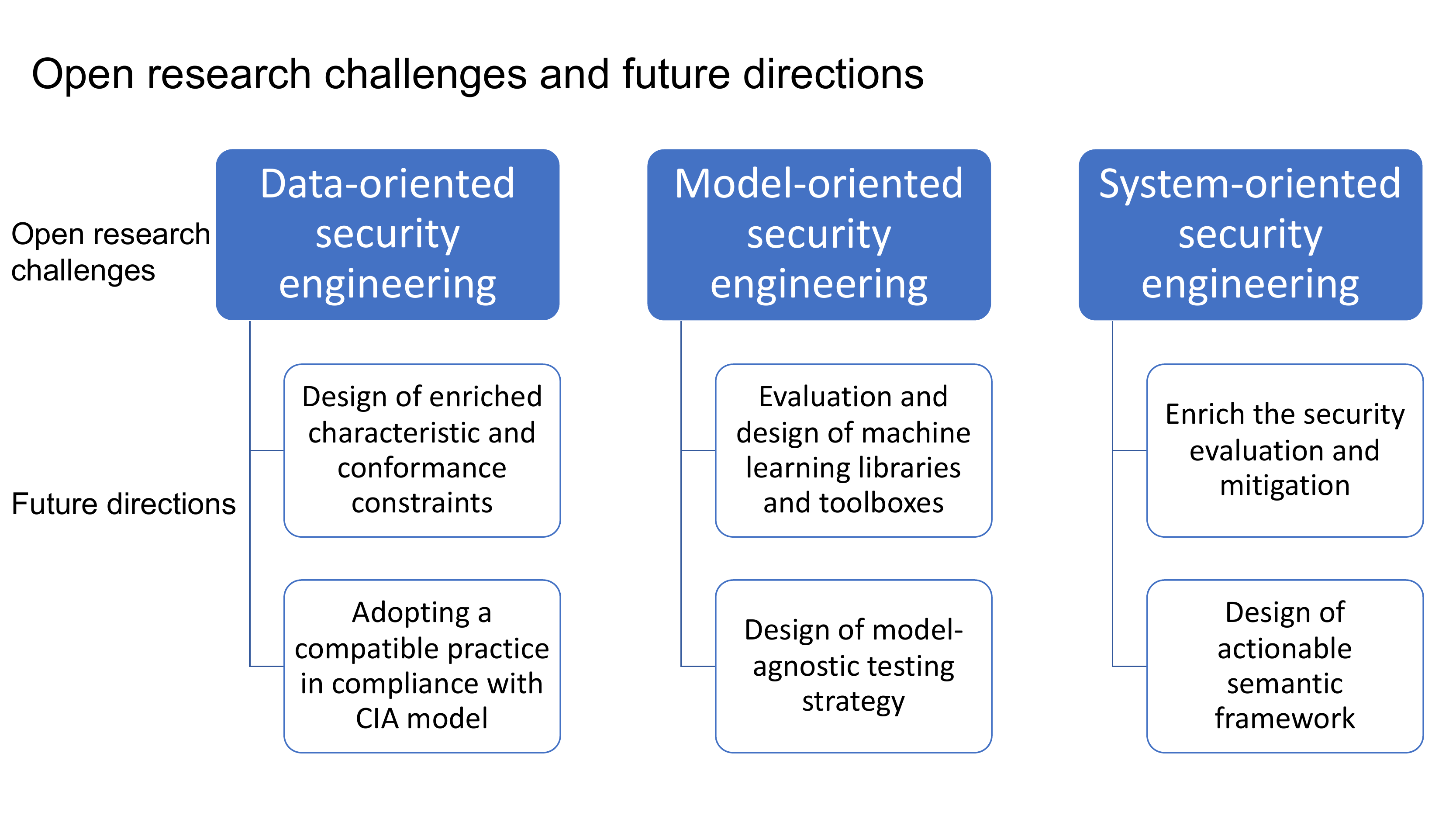}
  \caption{List of challenges and future directions for security engineering of MLBSS.}
  \label{fig:challenges_future}
\end{figure}
\indent While current security engineering practices in MLBSS has leveraged a large amount of efforts and resources, it is evident that a wide range of challenges of security engineering remain unexploited. Thus, in this section, we discuss the challenges with the security engineering for MLBSS from different perspectives aiming at the data-related, model-related and system-level framework elements. Accordingly, the future directions are discussed to address the challenges.\par
\subsection{Data-oriented security engineering for MLBSS}
This section brings attentions to the security engineering for MLBSS in terms of the data-oriented attacks, particularly focusing on the data management stage (section~\ref{subsec:data_management}). Firstly, it is recognised that current security practices has not fully considered the data confidentiality and integrity since the lack of an insightful understanding towards the security aspect of data management. Secondly, there is little work for the relaxation of data-oriented attacks in data management stage, which could provide a fundamental protocol (section~\ref{subsec:data_attacks}). To relax these two challenges, design of enriched characteristic and conformance constraints, and adopting a compatible practice in compliance with CIA model for data management would be potential future directions.\par
\subsubsection{Design of enriched characteristic and conformance constraints for data}\par
Data remains the pivotal component for machine learning-based systems, such that the malfunctional behaviour of systems can largely reside in the data. When splitting data for the training and evaluation of system deviates from each other, the output will become unreliable and unsafe for safety-critical systems. Thus, the introduction of conformance characteristic for data management stage can greatly alleviate this concern. Furthermore, the conformance should also be applied to the unseen data, which will be used for post-deployment inference against poisoning and backdoor attack.\par
\indent It is inspiring to adopting a compatible data auditing practice to ensure the data integrity from the work of Zhang et al.~\cite{zhang2020towards}. In this work, it indicates that the benign and adversarial data generated by same existing methods could be statistically categorised based on common uncertainty patterns. Thus, the defence capability could be further achieved by devising deviated adversarial data that do not follow the patterns for training, against the adversarial attack. Moreover, in ~\cite{fariha2021conformance,fariha2021coco}, the data profiling is investigated for data-driven systems to discover the conformance constraints for a dataset, with one recent framework proposed in ~\cite{galhotra2021dataexposer} called DataExposer. Hence, future work could be to study how to accurately define the profile, enrich the characteristics and conformance constraints for data understanding, validation, integration and cleaning.\par
\subsubsection{Adopting a compatible practice in compliance with CIA model for data management}\par
In section~\ref{subsec:data_management}, we can see that a general focus of the current practice is on the adoption of data quality practice. The aim is to ensure the accurate and robust machine learning-based functionalities during the inference time. While the data management stage involves the training and testing data for subsequent model construction and system deployment, the data should be obtained in a rigorous and standardised manner, indicating that CIA triad can be a better option for data management stage.\par
\indent Furthermore, the circumstance will become worse when the data management task is largely outsourced. Zhao et al.~\cite{zhao2020distributed} has explored the provable data possession verification scheme which is generally effective in cloud computing. The effectiveness of the distributed machine learning oriented data integrity verification scheme demonstrates the importance of adopting a compatible practice in the context. Thus, we call for future work to investigate a broader and more effective verification scheme, i.e., blockchain~\cite{zhu2019controllable}, to ensure that CIA model is tightly withhold for the data.\par
\subsection{Model-oriented security engineering for MLBSS}
For model-oriented security, current understanding and practices are not sufficient to secure the machine learning-based system against the identified attack types. Firstly, the machine learning libraries and toolboxes testing is a relevant topic to test the models yet little attention has been paid to. Secondly, current security practices targeting model testing could not generalise well for different types of machine learning models. Herein, we highlight following two future directions to tackle the challenges -- the evaluation and design of machine learning libraries and toolboxes; the design of model-agnostic system testing strategy.\par
\subsubsection{Evaluation and design of the machine learning libraries and toolboxes:}\par
Besides the training and testing data, there are several complex integral components whilst building MLBSS (see Fig.~\ref{fig:mlbss_process} and Fig.~\ref{fig:mlbss_component}). Currently, testing the robustness of the trained models has attracted most attentions to guarantee the quality and security of machine learning models. The most recent work ~\cite{wang2020deep} (see section.~\ref{subsec:model_system}) has targeted on another venue, which is to test the deep learning libraries including TensorFlow, Theano, CNTK and MXNet. A rather much more complicated and critical machine learning development community in comparison to traditional software development has been presented in this work.\par
\indent In the meantime, the various machine learning libraries and toolboxes could result in unknown security threats and uncertain performance impact for the machine learning model construction and system deployment stages. It is clear that testing the libraries and toolboxes would not be sufficient. Thus, a comprehensive evaluation for the machine learning libraries and toolboxes could be a more detailed option to alleviate the concern. Interestingly, a recent work from~\cite{kiraly2021designing} has initiated the discussion of a summary of best practices and a collection of machine learning toolboxes design patterns. It also highlights that it is promising to further extend the design of machine learning libraries and toolboxes following a certain, secure and practical design principles and patterns.\par
\subsubsection{Design of model-agnostic testing strategy for MLBSS:}\par
Although model testing has been investigated to achieve the testing adequacy with standards for MLBSS, it has been a fundamental challenge to test the learnt model and system against all potential adversarial input~\cite{du2019deepstellar,xie2019diffchaser,riccio2020testing} (see section.~\ref{subsec:model_system}).\par
\indent Generally, the testing strategy is devised to secure models against adversarial inputs. Recent work from Fan et al.~\cite{fan2021text} has extended the quality testing for recurrent neural network to backdoor attacks. However, it could be challenging to adopt the available testing strategy for GAN model, such as against membership inference attacks~\cite{chen2020gan}. Thus, a model-agnostic framework and benchmark of testing techniques for the evaluation of diverse attack types against different machine learning models will be much more valuable to collectively support model-oriented security engineering, as a future work.\par
\subsection{System-oriented security engineering for MLBSS}
The machine learning-based software system is essentially a software product following the identified pipelines (Fig.~\ref{fig:mlbss_process}). However, little work has been identified to provide the system-oriented security engineering to: 1) establish a comprehensive understanding and translation of problem definition (see section.~\ref{subsec:problrm_definition}), 2) enrich the safety evaluation of such systems (see section.~\ref{subsec:model_system}, ~\ref{subsec:system_maintenance}). Thus, we discuss two separate future directions to resolve the challenges: 1) enrich the security evaluation and mitigation methods for MLBSS; 2) design of actionable semantic framework, to support system-oriented security engineering for MLBSS.\par
\subsubsection{Enrich the security evaluation and mitigation for MLBSS}\par
Current security evaluation and mitigation regarding attack and defence methods are mostly conducted on either separate model or over single-node the cloud services, i.e., machine-learning-as-a-service from AWS and Google (see section.~\ref{subsec:model_attacks}, ~\ref{subsec:model_system}). Since the machine learning-based software systems could not be fully delivered with only hardware or software stack, how to consider the safety evaluation from both stacks could be more practical but is also much more challenging. Especially, the mitigation methods for hardware and software are technically different.\par
\indent While the research works for different artefacts in MLBSS remain isolated and require further investigation, it is appreciated to witness the academic efforts of applying hardware and software co-design mechanism to ensure the security and privacy for on-device machine learning~\cite{regazzoni2020machine,zhang2021democratic}. The enrichment of the security evaluation and mitigation covering hardware and software stacks to incorporate security-by-design and other security engineering processes and tools will be of critical importance for MLBSS from the system-level view.\par
\subsubsection{Design of actionable semantic framework to support security engineering for MLBSS}\par 
Recent works from ~\cite{peng2020first} and ~\cite{morley2021towards} have reported the practical usage of machine learning models in safety-critical domains, such as the healthcare and autonomous driving systems. Other works have considered the inclusion of \textit{hazard contribution modes}~\cite{smith2020hazard}, \textit{formal verification}~\cite{sun2019formal}, \textit{generalisation error of model}~\cite{zhao2020safety}, \textit{system accountability ontology}~\cite{naja2021semantic} for MLBSS. While numerous efforts have been dedicated to support the MLBSS security, it is recognised that more efforts are needed.\par
\indent In light of the early discussion by ~\cite{best2007model} for distributed information systems, we have seen a shift of distilling the security engineering practice with actionable framework design. Some recent works include the work from Ruiz et al.~\cite{ruiz2017security} for secure IT systems, the \textit{ten security principles} in ~\cite{mcgraw2020architectural} and the recent \textit{meta model} work by ~\cite{xiong2021towards} for machine learning-based systems. In this regard, it is anticipated that future work can incorporate more meaningful and actionable semantic framework and modelling language to support the system-oriented security engineering for MLBSS, based on the identified attack modes and security practices in this paper.\par

\section{Conclusion}
\label{sec:conclusion}
Security of machine learning-based software systems is a burgeoning topic for both researchers and industry practitioners in safety-critical systems (e.g., the cyber-physical system and intrusion detection systems). The various attack modes in terms of different artefacts, i.e., the data, model, hardware and system, raise huge concerns for security development of systems. While the system security engineering for MLBSS has not been explored in a comprehensive manner, this paper investigates the state-of-the-practice of system security for the machine learning-based software systems. It is noted that, based on the surveyed works, more efforts should be shifted from the particular model and algorithms side towards the security from a holistic view, such as the discussed security-by-design for MLBSS. Although there are several works pitching towards this direction, challenges such as the data-oriented, model-oriented, system-oriented security engineering are still open for improvements.\par
\section*{Acknowledgment}
The work has been supported by the Cyber Security Research Centre Limited whose activities are partially funded by the Australian Government’s Cooperative Research Centres Programme\par
\bibliographystyle{ACM-Reference-Format}
\bibliography{SurveySecurityMLBSS}
\end{document}